\documentclass[aoas,MSNbibl,nameyear,rotating,dvips]{arximspdf}
\usepackage{dcolumn}
\usepackage{graphicx}

\doi{10.1214/13-AOAS637} 
\volume{7}
\issue{3}
\pubyear{2013}
\firstpage{1474}
\lastpage{1496}

\makeatletter
\newcolumntype{d}[1]{D{.}{.}{#1}}
\newtheorem{theorem}{Theorem}
\makeatother

\begin{document}
\begin{frontmatter}

\title{Estimating restricted mean job tenures in semi-competing risk
data compensating victims of~discrimination}
\runtitle{Estimating restricted mean job tenures}

\begin{aug}
\author[A]{\fnms{Qing} \snm{Pan}\corref{}\ead[label=e1]{qpan@gwu.edu}}
\and
\author[A]{\fnms{Joseph L.} \snm{Gastwirth}\ead[label=e2]{jlgast@gwu.edu}}
\runauthor{Q. Pan and J. L. Gastwirth}
\affiliation{George Washington University}
\address[A]{Department of Statistics\\
George Washington Universtiy\\
Washington, DC 20052\\
USA\\
\printead{e1}\\
\phantom{E-mail:\ }\printead*{e2}}
%
\end{aug}

\received{\smonth{7} \syear{2012}}
\revised{\smonth{2} \syear{2013}}

%
\begin{abstract}
When plaintiffs prevail in a discrimination case, a major component of
the calculation of economic loss is the length of time they would have
been in the higher position had they been treated fairly during the
period in which the employer practiced discrimination. This problem is
complicated by the fact that one's eligibility for promotion is subject
to termination by retirement and both the promotion and retirement
processes may be affected by discriminatory practices. This
semi-competing risk setup is decomposed into a retirement process and a
promotion process among the employees. Predictions for the purpose of
compensation are made by utilizing the expected promotion and
retirement probabilities of similarly qualified members of the
nondiscriminated group. The restricted mean durations of three periods
are estimated---the time an employee would be at the lower position, at
the higher level and in retirement. The asymptotic properties of the
estimators are presented and examined through simulation studies. The
proposed restricted mean job duration estimators are shown to be robust
in the presence of an independent frailty term. Data from the reverse
discrimination case, \textit{Alexander v. Milwaukee}, where White-male
lieutenants were discriminated in promotion to captain are reanalyzed.
While the appellate court upheld liability, it reversed the original
damage calculations, which heavily depended on the time a plaintiff
would have been in each position. The results obtained by the proposed
method are compared to those made at the first trial. Substantial
differences in both directions are observed.
\end{abstract}

%

\end{frontmatter}

\section{Introduction}\label{sec1}
When plaintiffs prevail in a case involving hiring or promotion
discrimination, courts need to estimate the economic loss (compensatory
damages) they suffered. The problem arose in the reverse discrimination
case \textit{Alexander v. Milwaukee} (474 F. 3d 437, 7th Cir. 2007).
The City of Milwaukee and its Board of Fireman and Police Commissioners
were sued by seventeen Caucasian male members of the Police department
for discrimination with respect to promotions from lieutenant to
captain made during the seven years of Chief Jones' tenure. At the
trial, the jury found the defendants guilty of discrimination and
ordered compensatory damages for each plaintiff. On appeal, the Seventh
Circuit affirmed the district court's judgement with respect to
liability, but reversed the amount of damages awarded and remanded the
case for recalculation using more accurate estimates of the plaintiffs'
``lost chances.'' When there are more candidates than positions at the
time one or several openings are available, the economic damages need
to reflect the diminished probability of promotion that each plaintiff
lost (\textit{Bishop v. Gainer}, 272 F.3d 1009, 1015-16, 7th Cir. 2001;
\textit{Doll v. Brown}, 75 F.3d 1200, 1205-07, 1996; \textit{Griffin v.
Michigan Department of Corrections}, 5 F.3d 186, 189, 6th Cir. 1993).
Lost chances are the differences between the hypothetical promotion
probabilities the White-male plaintiffs would have received absent of
discrimination and the observed promotion probabilities.

Nonparametric [\citet{TabSta09}] and Bayesian [\citet{KadWoo04}; \citet{WooKad10}] methods have been
proposed to model employment data to determine liability. Their focus
is the accurate estimation of the regression coefficient for the
variable indicating the protected group membership. Our goal, however,
is to make predictions of compensatory damages, which requires
estimating the probabilities of having been promoted over time. \citet{PanGas09} used accelerated failure time models to estimate
hypothetical job tenures of plaintiffs who were discriminated in
hiring. Cox proportional hazards models [Cox (\citeyear{Cox72,Cox75})] are utilized
in this manuscript to predict promotion probabilities. Semiparametric
models are superior in several aspects. The models are more flexible,
as they only assume multiplicative covariate effects on the hazard
function without forcing the baseline to follow any specific
distribution. Second, when the time axis is set as calendar time, the
risk set at each promotion is composed of the eligible candidates at
that time. Therefore, the model is fitted by maximizing the
probabilities of selecting the individuals who actually received
promotion from the candidate pool. In this setting, the characteristics
of the cohort at risk for promotion are fairly stable because new
lieutenants replenish the group eligible for promotion and the
seniority of the remaining lieutenants increases when senior members retire.

Another complication arises because discrimination in one aspect of
employment may also affect other employment decisions. For example, the
same supervisors would select employees for promotion or reduction in
force. However, discrimination in layoffs is unlikely to occur in civil
service jobs which are protected by a tenure-type system, especially
when most incumbents have a fair amount of seniority. In the motivating
case \textit{Alexander v. Milwaukee}, the only form of termination is
retirement and there were no claims against discrimination in processes
other than promotion. Still, it is possible that retirement
is affected by the discriminatory practice in promotions. For example,
individuals with delayed or denied promotions due to discrimination
might have less incentives to remain employed after they become
eligible for retirement. Therefore, disparities in both the promotion
and retirement processes need to be considered when estimating the
economic damages. Promotion and retirement are semi-competing risks
[\citet{FinJiaCha01}], as the occurrence of retirement
terminates the promotion process, but retirements can be observed after
promotions. \citet{XuKalTai10} model such data structures
with three hazard/transition functions---the transition from no-event
status to the promoted status given that neither promotion nor
retirement happened previously, the transition from no-event status to
retired status given that neither events occurred, and the transition
from the promoted status to retired status given the promotion time. In
\textit{Alexander v. Milwaukee}, the number of years since becoming
eligible for retirement is the most important factor affecting
retirement hazard. Conditional on the number of years since becoming
eligible, the retirement pattern among lieutenants and that among
captains are similar [\textit{The New York Times}, Goldstein (\citeyear{Gol})].
Thus, the retirement process is modeled regardless of the promotion
status. Unlike competing risk data, the distribution of the retirement
times is identifiable [\citet{PenFin07}]. Furthermore, we model the
promotion risk conditional on that retirement has not occurred. Similar
to the cause-specific instantaneous rate of occurrence in competing
risks [\citet{Lin97}] which requires that no event of any type has
happened, the promotion hazard among nonretired employees is
conditional on that neither promotion nor retirement has occurred. It
is different from the promotion hazard conditional on that promotion
has not occurred. In summary, we model two hazard/transition functions
- the retirement process and the promotion process conditional on that
retirement has not occurred, respectively. This can be viewed as the
univariate event counterpart of \citet{CooLaw97} approach
modeling recurrent events in the presence of a terminal event, where
the recurrent event mean/rate at time point $t$ is conditional on that
the terminal event time is larger than $t$.

As the promotion and retirement probabilities are time-varying, a more
concise summary of the processes is provided by the durations of time
the plaintiff would be in each position. In biomedical studies,
researchers are interested in the restricted mean lifetime during a
specific period following the treatment, for example, 10 years, because
the treatment effects will not last beyond 10 years. \citet{CheTsi01}
used the difference in restricted mean lifetimes as a measure of
treatment effects. \citet{ZhaSch11} studied restricted mean
lifetime subject to informative censoring. In the motivating example,
lieutenants, captains and retired officers received different salaries
or pensions. Therefore, we estimate the hypothetical restricted mean
durations of the three statuses (not promoted, promoted, retired)
during the period for which the plaintiffs deserve compensation, that
is, from the beginning of the discriminatory practice to the cutoff
time determined by the court.

This manuscript studies the estimation of hypothetical restricted mean
job durations assuming that the discrimination did not happen for the
purpose of compensating prevailing plaintiffs. The paper is organized
as follows. The restricted mean duration estimator is constructed in
Section \ref{sec2}. The asymptotic distribution is derived in Section \ref{sec3}. The
accuracy of the estimator's asymptotic properties in moderate size
samples is examined by simulations in Section \ref{sec4}. The proposed estimator
is applied to the motivating reverse discrimination case in Section \ref{sec5}.
Additional issues in estimating compensation are discussed in Section \ref{sec6}.

\section{Proposed estimators}\label{sec2}
The notation will be described in the context of the motivating
example. There are $n$ employees indexed by $i$, $i=1,\ldots,n$. The
time axis is calendar time. When the $i${th} employee reaches the rank
of lieutenant, he or she becomes eligible for promotion to captain.
Thus, the calendar date an officer becomes lieutenant is the date that
individual enters the promotion process, denoted by $P_{1i}$. The date,
if any, when the $i$th employee is promoted to captain is denoted by
$P_{2i}^*$. Retirement terminates the promotion process and the data
collection time censors both promotions and retirements. Denote the
retirement date and the end of data collection for subject $i$ as
$R^*_{2i}$ and $C_i$, respectively. We observe $P_{2i}=\min\{P_{2i}^*,
R_{2i}^*, C_{i}\}$ and use $\delta^P_{i}=I(P_{2i}=P_{2i}^*)$ to
indicate an observed promotion event. Here $I(A)$ is an indicator
function which equals 1 when $A$ is true and 0 otherwise. Similar
notation is used for the retirement process. Police officers become
eligible for retirement after 25 years of service in the department.
Thus, they enter the retirement process on that day, $R_{1i}$. The end
of the retirement process, $R_{2i}^*$, occurs when the policeman
retires but is also censored at the time of data collection, $C_{i}$.
Thus, the observed time of the retirement process is $R_{2i}=\min\{
R_{2i}^*, C_{i}\}$, and $\delta^R_{i}=I(C_{i}>R_{2i}^*)$ indicates
whether $R_{2i}$ is a retirement time or censoring time. The observed
data consist of $n$ independent and identically distributed vectors,
$(P_{1i},P_{2i},\delta^P_{i},X'_i(t),R_{1i},R_{2i},\delta
^R_{i},Z'_i(t))'$, where the covariates for the promotion process,
$X_i(t)$, and those for the retirement process, $Z_i(t)$, include both fixed
and time-varying elements. The covariate sets $X'_i(t)$ and $Z'_i(t)$
often overlap, as some factors may play a role in both promotion and
retirement decisions.

The semi-competing risk data is decomposed into the retirement process
and the promotion process conditional on that retirement has not
occurred. The corresponding hazard functions are defined as follows:
\begin{eqnarray*}
d\Lambda_i^P(t)&=&\lim_{\triangle t\rightarrow0}
\frac{\operatorname{Pr}(t\leq
P^*_{2i}<t+\triangle t|P^*_{2i}\geq t,R^*_{2i}\geq t)}{\triangle
t}I(P_{1i}\leq t),
\\
d\Lambda_i^R(t)&=&\lim_{\triangle t\rightarrow0}
\frac{\operatorname{Pr}(t\leq
R^*_{2i}<t+\triangle t|R^*_{2i}\geq t)}{\triangle t}I(R_{1i}\leq t).
\end{eqnarray*}
Notice we require $P_{1i}\leq t$ and $R_{1i}\leq t$ so that $d\Lambda
_i^P(t)=0$ for $t<P_{1i}$ and $d\Lambda_i^R(t)=0$ for $t<R_{1i}$. The
two hazard functions are modeled by two Cox proportional hazards models
%
%
\begin{eqnarray}
d\Lambda_i^P(t)&=&d\Lambda_0^P(t)
\exp \bigl(X_i(t)'\beta_0 \bigr),
\label{PHP}
\\
d\Lambda_i^R(t)&=&d\Lambda_0^R(t)
\exp \bigl(Z_i(t)'\theta_0 \bigr),
\label{PHL}
\end{eqnarray}
where $d\Lambda_0^P(t)$ and $d\Lambda_0^R(t)$ denote the unspecified baseline
hazard functions over calendar time and the two vectors $\beta_0$,
$\theta_0$ represent the true values
of the regression coefficients. In the risk set of model (\ref{PHP}),
we do not differentiate lieutenants who are eligible for retirement and
those who are not yet eligible because seniority is a covariate and
officers with the same seniority usually have similar promotion
chances, regardless if eligible for retirement or not.

We assume
\begin{eqnarray*}
&&\lim_{\triangle t\rightarrow0}{\triangle t}^{-1}\operatorname{Pr} \bigl(t\leq
P_{2i}^* <t+\triangle t|t\leq P_{2i}^*, t\leq
R_{2i}^*, X_i(t),Z_i(t),R_{2i}^*
\bigr)
\\
&&\qquad=\lim_{\triangle t\rightarrow0}{\triangle t}^{-1}\operatorname{Pr} \bigl(t\leq
P_{2i}^* <t+\triangle t|t\leq P_{2i}^*, t\leq
R_{2i}^*, X_i(t),Z_i(t) \bigr),
\\
&&\lim_{\triangle t\rightarrow0}{\triangle t}^{-1}\operatorname{Pr} \bigl(t\leq
R_{2i}^* <t+\triangle t|t\leq R_{2i}^*, X_i(t),Z_i(t),P_{2i}^*
\bigr)
\\
&&\qquad=\lim_{\triangle t\rightarrow0}{\triangle t}^{-1}\operatorname{Pr} \bigl(t\leq
R_{2i}^* <t+\triangle t|t\leq R_{2i}^*, X_i(t),Z_i(t)
\bigr).
\end{eqnarray*}
That is, conditional on $X_i(t)$ and $Z_i(t)$, $d\Lambda_i^P(t)$ and
$d\Lambda_i^R(t)$ are assumed to be independent. This assumption would
be violated in the presence of latent variables affecting both
processes. However, models assuming latent variables would not be
accepted in courts (\textit{King v. Acosta Sales and Marketing Inc.},
678 F.3d 470 2012). Nevertheless, the sensitivity of the proposed
estimators to moderate deviations from this independence assumption is
studied in Table \ref{tab2}.

For any time point after $P_{1i}$, there are three possible statuses
for each subject---not retired and remaining a lieutenant, not retired
and being a captain, and retired. The probabilities for each of the
three statuses are
\begin{eqnarray*}
\operatorname{Pr} \bigl(P_{2i}^*>t, R_{2i}^*>t \bigr)&=&\operatorname{Pr}
\bigl(P_{2i}^*>t|R_{2i}^*>t \bigr)\operatorname{Pr} \bigl(R_{2i}^*>t
\bigr),
\\
\operatorname{Pr} \bigl(P_{2i}^*\leq t, R_{2i}^*>t \bigr)&=&\operatorname{Pr}
\bigl(P_{2i}^*\leq t|R_{2i}^*>t \bigr)\operatorname{Pr} \bigl(R_{2i}^*>t
\bigr),
\\
\operatorname{Pr} \bigl(R_{1i}\leq R_{2i}^*\leq t \bigr)&=&\operatorname{Pr}
\bigl(R_{2i}^*\leq t \bigr).
\end{eqnarray*}
Again, we do not differentiate between $t<R_{1i}$ and $t\geq R_{1i}$ in
the first two statuses. Therefore, the probability $\operatorname{Pr}(P_{2i}^*>t,
R_{2i}^*>t)$ is the probability of being a lieutenant and not retired
where the officer is either eligible or ineligible for retirement, and
$\operatorname{Pr}(P_{2i}^*\leq t, R_{2i}^*>t)$ is the probability of being a captain
and having not retired regardless of eligibility for retirement.

Notice that the probability of having been promoted to captain equals
the sum of the probability of being a captain and not retired and the
probability of having been promoted and retired. That is,
\[
\operatorname{Pr} \bigl(P_{2i}^*\leq t \bigr)=\operatorname{Pr} \bigl(P_{2i}^*\leq t,
R_{2i}^*>t \bigr)+\operatorname{Pr} \bigl(P_{2i}^*\leq t, R_{2i}^*
\leq t \bigr)
\]
for any $t > P_{1i}$. Similarly,
\begin{eqnarray*}
\operatorname{Pr} \bigl(P_{2i}^*> t \bigr)&=&\operatorname{Pr} \bigl(P_{2i}^*> t,
R_{2i}^*>t \bigr)+\operatorname{Pr} \bigl(P_{2i}^*> t, R_{2i}^*
\leq t \bigr),
\\
\operatorname{Pr} \bigl(R_{2i}^*\leq t \bigr)&=&\operatorname{Pr} \bigl(R_{2i}^*\leq t,
P_{2i}^*>t \bigr)+\operatorname{Pr} \bigl(R_{2i}^*\leq t, P_{2i}^*
\leq t \bigr),
\\
\operatorname{Pr} \bigl(R_{2i}^*> t \bigr)&=&\operatorname{Pr} \bigl(R_{2i}^*> t,
P_{2i}^*>t \bigr)+\operatorname{Pr} \bigl(R_{2i}^*> t, P_{2i}^*
\leq t \bigr).
\end{eqnarray*}
The three probabilities of interests, $\operatorname{Pr}(P_{2i}^*>t, R_{2i}^*>t)$,
$\operatorname{Pr}(P_{2i}^*\leq t, R_{2i}^*>t)$ and $\operatorname{Pr}(R_{2i}^*\leq t)$, can be
estimated from models (\ref{PHP}) and (\ref{PHL}). But $\operatorname{Pr}(P_{2i}^*\leq
t, R_{2i}^*\leq t)$ and $\operatorname{Pr}(P_{2i}^*> t, R_{2i}^*\leq t)$ are
unidentifiable because the distribution of $P_{2i}^*$ is unobserved
after $R_{2i}^*$ [\citet{FinJiaCha01}; \citet{XuKalTai10}].

The expected lengths of time being a lieutenant ($T^{\mathrm{lt}}$), a captain
($T^{\mathrm{cap}}$) or retired ($T^{\mathrm{rt}}$) restricted to $[\tau_0,\tau_1]$ are
%
%
\begin{eqnarray}\label{ETa}
E \bigl(T^{\mathrm{lt}} \bigr)&=&\int_{\max(\tau_0, P_{1i})}^{\tau_1} \operatorname{Pr}
\bigl(P_{2i}^*> t> P_{1i}, R_{2i}^*>t \bigr)\,dt
\nonumber
\\
&=&\int_{\max(\tau_0, P_{1i})}^{\tau_1} \operatorname{Pr} \bigl(P_{2i}^*> t>
P_{1i}|R_{2i}^*>t \bigr)\operatorname{Pr} \bigl(R_{2i}^*>t \bigr)
\,dt,
\nonumber
\\
E \bigl(T^{\mathrm{cap}} \bigr)&=&\int_{\max(\tau_0, P_{1i})}^{\tau_1}
\operatorname{Pr} \bigl(P_{2i}^*\leq t, R_{2i}^*>t \bigr)\,dt
\\
&=&\int_{\max(\tau_0, P_{1i})}^{\tau_1} \operatorname{Pr} \bigl(P_{2i}^*\leq
t|R_{2i}^*>t \bigr)\operatorname{Pr} \bigl(R_{2i}^*>t \bigr) \,dt,
\nonumber
\\
E \bigl(T^{\mathrm{rt}} \bigr)&=&\int_{\max(\tau_0, P_{1i})}^{\tau_1}
\operatorname{Pr} \bigl(R_{2}^*\leq t \bigr)\,dt.\nonumber
\end{eqnarray}
The restricted mean job duration calculation starts from $\tau_0$ or
$P_{1i}$, whichever happens later, because an officer could not be
discriminated against promotion to captain until $P_{1i}$ (date
becoming a lieutenant) and $\tau_0$ (the time the discriminatory chief
was appointed).

The number of observed promotion and retirement events up to time point~$t$ are defined as $N_i^P(t)=\delta^P_{i}I(P_{2i}\leq t)$ and
$N_i^R(t)=\delta^R_{i}I(R_{2i}\leq t)$, respectively. Let
$dN_i^P(s)=N_i^P(s)-N_i^P(s^-)$ and $dN_i^R(s)=N_i^R(s)-N_i^R(s^-)$,
then $N_i^P(t)=\int_{\tau_0}^t \,dN_i^P(s)$ and $N_i^R(t)=\int_{\tau_0}^t
\,dN_i^R(s)$. The corresponding at-risk indicators are denoted by
$Y_i^P(t)=I(P_{1i}\leq t\leq P_{2i})$ and $Y_i^R(t)=I(R_{1i}\leq t\leq
R_{2i})$. The parameters $\beta_0$ and $\theta_0$ are estimated by
$\widehat{\beta}$ and $\widehat{\theta}$, the solution to the partial
likelihood score
functions, $U^P(\beta)=0$ and $U^R(\theta)=0$, which are defined as
\begin{eqnarray*}
U^P(\beta) &=&\sum_{i=1}^n
\int_{\tau_0}^{\tau_1}U_i^P(\beta)=
\sum_{i=1}^n\int_{\tau_0}^{\tau_1}
\bigl\{X_i(t)-\overline{X}(t;\beta) \bigr\} \,dN_i^P(t),
\\
U^R(\theta)&=&\sum_{i=1}^n
\int_{\tau_0}^{\tau_1}U_i^R(\beta)=
\sum_{i=1}^n\int_{\tau_0}^{\tau_1}
\bigl\{Z_i(t)-\overline{Z}(t;\theta) \bigr\} \,dN_i^R(t),
\end{eqnarray*}
where
\begin{eqnarray*}
\overline{X}(\beta;t)&=&\frac{S_P^{(1)}(t;\beta)} {S_P^{(0)}(t;\beta
)},
\\
\overline{Z}(\theta;t)&=&\frac{S_R^{(1)}(t;\theta)}{S_R^{(0)}(t;\theta
)},
\\
S_P^{(k)}(t;\beta)&=&n^{-1}\sum
_{i=1}^nY^P_i(t)e^{\beta'
X_i(t)}X_i(t)^{\otimes{k}},
\\
S_R^{(k)}(t;\theta)&=&n^{-1}\sum
_{i=1}^nY^R_i(t)e^{\theta
'Z_i(t)}Z_i(t)^{\otimes{k}},\qquad k=0,1,2,
\end{eqnarray*}
where $X_i(t)^{\otimes{0}}=1, Z_i(t)^{\otimes{0}}=1$, $X_i(t)^{\otimes
{1}}=X_i(t),Z_i(t)^{\otimes{1}}=Z_i(t)$ and $X_i^{\otimes
{2}}=X_i(t)X_i(t)',Z_i^{\otimes{2}}=Z_i(t)Z_i(t)'$. The Breslow--Aalen
baseline hazard estimators, $\widehat{\Lambda}_0^P(t;\widehat{\beta})$
and $\widehat{\Lambda}_0^R(t;\widehat{\theta})$, are employed, where
\begin{eqnarray*}
d\widehat{\Lambda}^P_0(t;\beta) &=&n^{-1}\sum
_{i=1}^nS_P^{(0)}(t;
\beta)^{-1} \,dN^P_i(t),
\\
d\widehat{\Lambda}^R_0(t;\theta)&=&n^{-1}\sum
_{i=1}^nS_R^{(0)}(t;
\theta)^{-1}\,dN^R_i(t).
\end{eqnarray*}

The covariates $({X}_i(t),{Z}_i(t))$ can be decomposed into
$({X}_i(t)=\{x_{i1}, X_{i2}(t)\},\break{Z}_i(t)=\{z_{i1},Z_{i2}(t)\})$, where
$x_{i1}=z_{i1}$ are indicators for White-male and
$X_{i2}(t),  Z_{i2}(t)$ denote other factors considered in the promotion
and retirement processes. We set
\begin{eqnarray*}
\tilde{X}_i(t)&=& \bigl\{0,X_{i2}(t) \bigr\},
\\
\tilde{Z}_i(t)&=& \bigl\{0,Z_{i2}(t) \bigr\}.
\end{eqnarray*}
That is, the White-male indicator is set to zero while other covariates
$(X_{i2}(t),\break Z_{i2}(t))$ remain the same. The goal is to estimate the
promotion/retirement probabilities and restricted mean job tenures the
plaintiffs would have received had they been treated the same as the
nondiscriminated members of the department. This is achieved by using
the hypothetical covariate values $(\tilde{X}_i(t),\tilde{Z}_i(t))$ in
estimating $E(T^{\mathrm{cap}})$, $E(T^{\mathrm{lt}})$ and $E(T^{\mathrm{rt}})$.

The survival functions from models (\ref{PHP}) and (\ref{PHL}) for a
subject with covariate values $(\tilde{X}_i(t),\tilde{Z}_i(t))$ and
entry times $P_{1i}, R_{1i}$ are estimated by
%
%
\begin{eqnarray}
\label{subjectS} \widehat{S}^P \bigl(t|\tilde{X}_i(t)
\bigr)&=&\exp \biggl\{-\int_{P_{1i}}^t \,d\widehat{
\Lambda }^P_0(r)e^{\widehat{\beta}'\tilde{X}_i(r)} \biggr\},
\nonumber
\\[-8pt]
\\[-8pt]
\nonumber
\widehat{S}^R \bigl(t|\tilde{Z}_i(t) \bigr)&=&\exp \biggl
\{- \int_{R_{1i}}^t \,d\widehat{\Lambda}^R_0(r)e^{\widehat{\theta}'\tilde{Z}_i(r)}
\biggr\},
\end{eqnarray}
where unknown parameters $d\Lambda^P_0(t), d\Lambda^R_0(t),\beta, \theta
$ are replaced by their maximum partial likelihood estimators.
Furthermore, combining (\ref{ETa}) and (\ref{subjectS}),
%
%
\begin{eqnarray}
\label{prediction}
&&\widehat{E} \bigl(T^{\mathrm{lt}}|\tilde{X}_i(t),
\tilde{Z}_i(t) \bigr)\nonumber\\
&&\qquad=\int_{\max(\tau_0,
P_{1i})}^{\tau_1}
\widehat{S}^P \bigl(t|\tilde{X}_i(t) \bigr)
\widehat{S}^R \bigl(t|\tilde{Z}_i(t) \bigr)\,dt
\\
&&\qquad=\int_{\max(\tau_0, P_{1i})}^{\tau_1}\exp \biggl\{-\int
_{P_{1i}}^t \,d\widehat{\Lambda}^P_0(u)e^{\widehat{\beta}'\tilde{X}_i(u)}
\biggr\}\nonumber\\
&&\qquad\quad\hspace*{46pt}{}\times\exp \biggl\{-\int_{R_{1i}}^t \,d\widehat{
\Lambda}^R_0(u)e^{\widehat{\theta}'\tilde
{Z}_i(u)} \biggr\}\,dt,\nonumber
\\
\label{prediction}
&&\widehat{E} \bigl(T^{\mathrm{cap}}|\tilde{X}_i(t),
\tilde{Z}_i(t) \bigr)\nonumber\\
&&\qquad=\int_{\max(\tau_0,
P_{1i})}^{\tau_1}
\bigl\{1-\widehat{S}^P \bigl(t|\tilde{X}_i(t) \bigr)
\bigr\} \widehat{S}^R \bigl(t|\tilde{Z}_i(t) \bigr)\,dt
\nonumber
\\[-8pt]
\\[-8pt]
\nonumber
&&\qquad=\int_{\max(\tau_0, P_{1i})}^{\tau_1} \biggl[1-\exp \biggl\{-\int
_{P_{1i}}^t \,d\widehat{\Lambda}^P_0(u)e^{\widehat{\beta}'\tilde{X}_i(u)}
\biggr\} \biggr]\\
&&\qquad\quad\hspace*{46pt}{}\times\exp \biggl\{ -\int_{R_{1i}}^t d
\widehat{\Lambda}^R_0(u)e^{\widehat{\theta}'\tilde
{Z}_i(u)} \biggr\}\,dt,\nonumber
\\
\label{prediction}
&&\widehat{E} \bigl(T^{\mathrm{rt}}|\tilde{X}_i(t),
\tilde{Z}_i(t) \bigr)\nonumber\\
&&\qquad=\int_{\max(\tau_0,
P_{1i})}^{\tau_1}
\bigl\{1-\widehat{S}^R \bigl(t|\tilde{Z}_i(t) \bigr)
\bigr\}\,dt
\\
&&\qquad=\int_{\max(\tau_0, P_{1i})}^{\tau_1} \biggl[1-\exp \biggl\{-\int
_{R_{1i}}^t \,d\widehat{\Lambda}^R_0(u)e^{\widehat{\theta}'\tilde{Z}_i(u)}
\biggr\} \biggr]\,dt.
\nonumber
\end{eqnarray}

Here, we use $\widehat{S}^P(t|Z_i(t))$ to estimate $\operatorname{Pr}(P_{2i}^{\ast}>t>P_{1i}|R_{2i}^{\ast}>t)$
because the two processes being modeled are assumed
to be independent given the covariates. Therefore, the probability of being a
lieutenant among all nonretired employees equal that of subjects who are not retired by time $t$.

\section{Asymptotic properties}\label{sec3}
The following regularity conditions are assumed:
\begin{longlist}[(a)]
\item[(a)] $(P_{1i},P_{2i},\delta^P_i,X'_i(t),R_{1i},R_{2i},\delta
^R_i,Z'_i(t))'$ are independent
and identically distributed;

\item[(b)] $\lim_{\triangle t\rightarrow0}{\triangle t}^{-1}\operatorname{Pr}(t\leq
P_{2i}^* <t+\triangle t|t\leq P_{2i}^*, t< R_{2i}^*,
t<C_i,X_i(t))=\break
\lim_{\triangle t\rightarrow0}{\triangle t}^{-1}\operatorname{Pr}(t\leq P_{2i}^*
<t+\triangle t|t\leq P_{2i}^*, t< R_{2i}^*, X_i(t))$;

\item[(c)] $\lim_{\triangle t\rightarrow0}{\triangle t}^{-1}\operatorname{Pr}(t\leq
{R_{2i}^*}<t+\triangle t|t\leq R_{2i}^*, t<C_i,Z_i(t))=\break
\lim_{\triangle t\rightarrow0}{\triangle t}^{-1}\operatorname{Pr}(t\leq
{R_{2i}^*}<t+\triangle t|t\leq R_{2i}^*, Z_i(t));$

\item[(d)] $\int_{\tau_0}^{\tau_1}\,d\Lambda_{0}^P(t)<\infty$,
$\int_{\tau_0}^{\tau_1}\,d\Lambda_{0}^R(t)<\infty$;

\item[(e)] Elements of $Z_{i}(t)$ and $X_{i}(t)$ are bounded almost
surely for $t\in[\tau_0,\tau_1]$;

\item[(f)] Positive-definiteness of the Hessian matrices, $A^P(\beta)$
and $A^R(\theta)$,
where
\begin{eqnarray*}
A^P(\beta)&=&E \biggl[\int_{\tau_0}^{\tau_1}
\bigl\{X_i(t)-\overline{X}(t;\beta) \bigr\}^{\otimes2}Y_i^P(t)e^{\beta
'X_i(t)}
\,d \widehat{\Lambda}^P_0(t) \biggr],
\nonumber
\\
A^R(\theta)&=&E \biggl[\int_{\tau_0}^{\tau_1}
\bigl\{Z_i(t)-\overline{Z}(t;\theta) \bigr\}^{\otimes2}Y_i^R(t)e^{\theta
'Z_i(t)}
\,d \widehat{\Lambda}^R_0(t) \biggr].\nonumber
\end{eqnarray*}
\end{longlist}
Condition (a) is usually satisfied unless there are clustered or
grouped subjects. Conditions (b) and (c) assume noninformative and
independent censoring. Condition (d) and (e) requires the cumulative
baseline hazard functions and the covariates to be bounded. Condition
(f) guarantees the Hessian matrices are nonsingular and their inverses exist.

The asymptotic properties for $\widehat{E}(T^{\mathrm{cap}}|\tilde{X}_i(t),\tilde
{Z}_i(t))$ are summarized in Theorems \ref{th1} and \ref{th2}. The details of the
proofs are provided in the \hyperref[app]{Appendix}.

\begin{theorem}\label{th1}
Under conditions \textup{(a)} to \textup{(f)},
\[
\widehat{E} \bigl(T^{\mathrm{cap}}|\tilde{X}_i(t),
\tilde{Z}_i(t) \bigr)\stackrel{a.s.} {\longrightarrow} E
\bigl(T^{\mathrm{cap}}|\tilde{X}_i(t),\tilde{Z}_i(t)
\bigr).
\]
\end{theorem}

\begin{theorem}\label{th2}
Under conditions \textup{(a)} to \textup{(f)},
$n^{{1}/{2}}\{\widehat{E}(T^{\mathrm{cap}}|\tilde{X}_i(t),\tilde
{Z}_i(t))- E(T^{\mathrm{cap}}|\tilde{X}_i(t),\tilde{Z}_i(t))\}$ converges weakly
to a mean-zero Gaussian process with variance $E[\{\int_{\tau_0}^{\tau
_1}\xi_i(t)\,dt\}^2]$, where
\begin{eqnarray*}
\xi_i(t)&=& \bigl\{\widehat{S}^R \bigl(t|
\tilde{Z}_i(t) \bigr) \bigl(-\xi_{i1}^P(t)-\xi
_{i2}^P(t) \bigr)\\
&&\hspace*{4pt}{}+ \bigl(1-S^P \bigl(t|
\tilde{X}_i(t) \bigr) \bigr) \bigl(\xi_{i1}^R(t)+
\xi_{i2}^R(t) \bigr) \bigr\},
\\
\xi^P_{i1}(t)&=&-\widehat{S}^P \bigl(t|
\tilde{X}_i(t) \bigr)\int_{P_{1i}}^t
\bigl[e^{\beta
_0'\tilde{X}_i(u)} \bigl\{\tilde{X}_i(u)-\overline{X}(u;
\beta_0) \bigr\}\,d\widehat{\Lambda}^P_0(u;
\beta_0) \bigr]\\
&&{}\times \bigl\{A^P(\beta_0) \bigr
\}^{-1}U_i^P(\beta_0),
\\
\xi_{i2}^P(t)&=&-S^P \bigl(t|
\tilde{X}_i(t) \bigr)\int_{P_{1i}}^te^{\beta'_0\tilde
{X}_{i}(u)}
\frac{dM_{i}^P(u;\beta_0)}{s^{(0)}_P(u;\beta_0)},
\\
\xi^R_{i1}(t)&=&-\widehat{S}^R \bigl(t|
\tilde{Z}_i(t) \bigr)\int_{R_{1i}}^t
\bigl[e^{\theta_0'\tilde{Z}_i(u)} \bigl\{\tilde{Z}_i(u)-\overline{Z}(u;
\theta_0) \bigr\}\,d\widehat{\Lambda}^R_0(u;
\theta_0) \bigr]\\
&&{}\times{A^R(\theta_0)}^{-1}U_i^R(
\theta_0),
\\
\xi_{i2}^R(t)&=&-S^R \bigl(t|
\tilde{Z}_i(t) \bigr)\int_{R_{1i}}^te^{\theta'_0\tilde
{Z}_{i}(u)}
\frac{dM_{i}^R(u;\theta_0)}{s^{(0)}_R(u;\theta_0)},
\nonumber
\\
s^{(0)}_P(u;\beta)&=&\lim_{n\rightarrow\infty}S^{(0)}_P(u;
\beta),
\\
dM_i^P(u;\beta)&=&dN_i^P(u)-Y_i^P(u)e^{\beta'X_i(u)}
\,d \Lambda_0^P(u),
\\
s^{(0)}_R(u;\theta)&=&\lim_{n\rightarrow\infty}S^{(0)}_R(u;
\theta),
\\
dM_i^R(u;\theta)&=&dN_i^R(u)-Y_i^R(u)e^{\theta'Z_i(u)}
\,d \Lambda_0^R(u).
\end{eqnarray*}
\end{theorem}

Here, $E[\{\int_{\tau_0}^{\tau_1}\xi_i(t)\,dt\}^2]$ can be estimated by
replacing parameters with their maximum likelihood estimates and
expectations with sample averages. The asymptotic distributions of
$\widehat{E}(T^{\mathrm{lt}}|\tilde{X}_i(t),\tilde{Z}_i(t))$ and $\widehat
{E}(T^{\mathrm{rt}}|\tilde{X}_i(t),\tilde{Z}_i(t))$ are derived similarly.

\section{Simulation studies}\label{sec4}
Data sets with $n=500$ independent and identically distributed pairs of
promotion among employees and retirement times are generated. Both
processes start at time zero for all subjects. The hazard functions
follow proportional hazards models,
\begin{eqnarray*}
d\Lambda^P(t|X_i)&=&d\Lambda_0^P(t)e^{\beta_1X_{i1}+\beta
_2X_{i2}},
\\
d\Lambda^R(t|Z_i)&=&d\Lambda_0^R(t)e^{\theta_1Z_{i1}+\theta
_2Z_{i2}},
\end{eqnarray*}
where $X_{i1}=Z_{i1}$ (minority or unfavored group indicator) is
distributed as {Bernoulli} $(0.5)$, and covariates $X_{i2}$ and $Z_{i2}$
follow Uniform $(0,10)$ and Normal $(0,4)$, respectively. Both baseline
functions are constant over time where $\lambda_0^P(t)=\frac{1}{10}$
and $\lambda_0^R(t)=\frac{1}{60}$. The coefficients for $X_{i1}$ are
$Z_{i1}$ are set to $(\beta_1=0, \theta_1=0)$, $(\beta_1=-0.5, \theta
_1=0.5)$ or $(\beta_1=-0.5, \theta_1=1)$. The coefficients for $X_{i2}$
and $Z_{i2}$ are $\beta_2=\theta_2=0.1$. Censoring is uniformly
distributed on $(0,200)$, which leads to approximately $36\%$ censoring
in the promotion process and $14\%$ in the retirement process. Each
data configuration is repeated 1000 times.

Table \ref{tab1} lists the performance of $\widehat{E}(T^{\mathrm{cap}}|\tilde
{X}_i(t),\tilde{Z}_i(t))$ restricted to $[0, \tau]$. Various
combinations of $\tilde{X}_{i}, \tilde{Z}_{i}$ and two time points $\tau
=5$ and $\tau=10$ are examined. In all these configurations, $\widehat
{E}(T^{\mathrm{cap}}|\tilde{X}_i(t),\tilde{Z}_i(t))$ is close to the true values
obtained by numerical integration, and the average estimated asymptotic
standard errors (ASE) agree with the empirical standard deviations
(ESD). The empirical coverage probabilities (CP) are close to the
nominal value of $0.95$.
%
%
\begin{table}
\caption{Simulation results: Performance of $\widehat{E}(T^{\mathrm{cap}}|\tilde
{X}_i(t),\tilde{Z}_i(t))$}\label{tab1}
\begin{tabular*}{\textwidth}{@{\extracolsep{\fill}}lccd{2.0}cd{2.3}ccc@{}}
\hline
\multicolumn{1}{@{}l}{$\bolds{\theta_1}$}&
\multicolumn{1}{c}{$\bolds{\tilde{X}_{i2}}$}&
\multicolumn{1}{c}{$\bolds{\tilde{Z}_{i2}}$}&
\multicolumn{1}{c}{$\bolds{\tau}$}&
\multicolumn{1}{c}{$\bolds{E(T^{\mathrm{cap}})}$}&
\multicolumn{1}{c}{\textbf{bias}}&
\multicolumn{1}{c}{\textbf{ESD}} & \multicolumn{1}{c}{\textbf{ASE}} &\multicolumn{1}{c@{}}{\textbf{CP}} \\
\hline
0 &1&1&5 &1.280&-0.001& 0.101& 0.096&0.941 \\
0 &1&1&10&3.784& 0.002& 0.237& 0.207&0.906 \\
0 &2&5&5 &1.344& 0.006& 0.117& 0.106&0.925 \\
0 &2&5&10&3.816&-0.012& 0.239& 0.222&0.920 \\
0.5&1&1&5 &1.280& 0.003& 0.098& 0.095&0.933 \\
0.5&1&1&10&3.784&-0.010& 0.233& 0.211&0.927 \\
0.5&2&5&5 &1.344&-0.004& 0.113& 0.104&0.929 \\
0.5&2&5&10&3.816& 0.002& 0.244& 0.226&0.925 \\
1 &1&1&5 &1.280&-0.003& 0.100& 0.097&0.940 \\
1 &1&1&10&3.784&-0.005& 0.225& 0.219&0.937 \\
1 &2&5&5 &1.344& 0.001& 0.112& 0.107&0.938 \\
1 &2&5&10&3.816&-0.010& 0.254& 0.236&0.935 \\
\hline
\end{tabular*}\vspace*{-3pt}
\end{table}

%
%
\begin{sidewaystable}
\tablewidth=\textwidth
\caption{Sensitivity of $\widehat{E}(T^{\mathrm{cap}}|\tilde{X}_i(t),\tilde
{Z}_i(t))$ to frailty. $\beta_0=-0.5, \theta_0=0.5$}\label{tab2}
\begin{tabular*}{\textwidth}{@{\extracolsep{\fill}}lcccccccccccd{2.3}ccc@{}}
\hline
&\multicolumn{4}{c}{$\bolds{\widehat{\beta}_1}$}&&\multicolumn{4}{c}{$\bolds{\widehat
{\theta}_1}$}&&\multicolumn{5}{c@{}}{$\bolds{\widehat{E}(T^{\mathrm{cap}})}$}\\[-6pt]
&\multicolumn{4}{c}{\hrulefill}&&\multicolumn{4}{c}{\hrulefill}&&\multicolumn{5}{c@{}}{\hrulefill}\\
\multicolumn{1}{@{}l}{\textbf{Var}}&\multicolumn{1}{c}{\textbf{bias}} & \multicolumn{1}{c}{\textbf{ESD}} &
\multicolumn{1}{c}{\textbf{ASE}} &\multicolumn{1}{c}{\textbf{CP}} &&\multicolumn{1}{c}{\textbf{bias}} &
\multicolumn{1}{c}{\textbf{ESD}} & \multicolumn{1}{c}{\textbf{ASE}} &\multicolumn{1}{c}{\textbf{CP}} &&
\multicolumn{1}{c}{\textbf{true}} &\multicolumn{1}{c}{\textbf{bias}} & \multicolumn{1}{c}{\textbf{ESD}} &
\multicolumn{1}{c}{\textbf{ASE}} &\multicolumn{1}{c@{}}{\textbf{CP}} \\
\hline
0.5&0.056&0.117&0.116&0.923&&$-0.125$&0.103&0.103&0.769&&1.164&-0.036&0.101&0.096&0.911
\\
0.5&0.057&0.117&0.116&0.923&&$-0.118$&0.105&0.103&0.782&&3.303&-0.091&0.223&0.208&0.908
\\
0.5&0.065&0.113&0.116&0.922&&$-0.124$&0.100&0.103&0.771&&1.151&0.017&0.110&0.107&0.936
\\
0.5&0.052&0.119&0.116&0.923&&$-0.122$&0.104&0.103&0.759&&3.160&0.109&0.253&0.230&0.908
\\[6pt]
1
&0.087&0.117&0.119&0.891&&$-0.187$&0.113&0.108&0.588&&1.071&-0.047&0.095&0.093&0.912
\\
1
&0.094&0.121&0.120&0.869&&$-0.185$&0.109&0.108&0.621&&2.950&-0.077&0.217&0.205&0.908
\\
1
&0.087&0.125&0.120&0.873&&$-0.189$&0.107&0.108&0.568&&1.050&0.015&0.113&0.104&0.931
\\
1
&0.087&0.121&0.119&0.882&&$-0.185$&0.109&0.108&0.575&&2.796&0.119&0.239&0.227&0.915
\\[6pt]
2
&0.103&0.128&0.129&0.871&&$-0.243$&0.118&0.118&0.448&&0.930&-0.029&0.094&0.090&0.916
\\
2
&0.105&0.129&0.128&0.855&&$-0.251$&0.118&0.118&0.433&&2.456&-0.014&0.210&0.201&0.928
\\
2
&0.107&0.130&0.128&0.862&&$-0.244$&0.118&0.118&0.448&&0.900&0.040&0.108&0.102&0.928
\\
2
&0.106&0.127&0.128&0.871&&$-0.252$&0.120&0.117&0.421&&2.295&0.098&0.235&0.225&0.864
\\
\hline
\end{tabular*}
\end{sidewaystable}

Promotion among employees and retirement are assumed to be independent
conditional on the covariates $X_i(t)$ and $Z_i(t)$. However,
there might be unmeasured latent variables that affect both processes
and lead to correlated residuals from the proportional hazards models
(\ref{PHP}) and (\ref{PHL}). To test the robustness of the proposed
restricted mean job duration estimators to unadjusted frailties,
sensitivity analyses are carried out. Following the literature [\citet{YeKalSch07}],
a gamma frailty is generated, which multiplies the hazard
rates of both processes. The frailty terms are independent from
$X_{i1}$ and $Z_{i1}$. Without loss of generality, we set the mean of
the gamma random variable to be one and examine three values for the
variance: 0.5, 1 and 2. The same parameter and covariate values used in
the middle four rows in Table \ref{tab1} are employed in the sensitivity
analysis. The true values of $E(T^{\mathrm{cap}})$ are obtained by integrating
over the gamma frailty distribution and listed in column 10 of Table \ref{tab2}.
The estimates of the regression coefficients from the two Cox
proportional hazards models ignoring the frailty term are biased toward
zero. The ASEs are close to the corresponding ESDs even in the presence
of a frailty. Although constructed from the biased coefficients and
baseline hazard estimates ignoring frailties, the restricted mean job
duration estimates are not very different from the true values, where
the magnitudes of most of the biases are less than 10$\%$ of the true
values. The proposed variance estimates of $\widehat{E}(T^{\mathrm{cap}})$ are
slightly smaller than the empirical ones, and the coverage
probabilities range from 86$\%$ to 94$\%$. In summary, the proposed
methods are reasonably robust in making predictions in the presence of
frailty terms. Unlike clinical trials, employees are not randomized
into the protected and unprotected groups, so a latent variable may be
distributed differently in the two groups. If such a confounder exists,
the coefficient estimators $\widehat{\beta}_1$, $\widehat{\theta}_1$
and the mean restricted job duration estimators will be biased (results
not shown).

\section{Application}\label{sec5}

Officer Arthur Jones was the Police Chief of the City of Milwaukee from
November 18th, 1996 to November 18th, 2003. During his tenure there
were 112 White-male lieutenants and 34 female or non-White lieutenants
who were eligible for promotion. He selected 21 White-males and 20
others for promotion, thus, White-males had a promotion rate of $19\%$
in contrast to the $59\%$ rate for females and non-White-males.
Furthermore, among promoted individuals, the average length of time the
White-male lieutenants served before becoming captain was 7.36 years,
while the average length for the others was 3.02 years (\mbox{$p$-value} of the
Wilcoxon test $< 0.001$). Seventeen White-male lieutenants brought a
reverse discrimination case against the City of Milwaukee. At trial, a
jury found the defendants liable for discrimination against the
plaintiffs in promotion to captain. Compensatory damages for the
plaintiffs' economic loss in wages and pensions as well as punitive
damages were ordered. The defendants' motions to vacate both the
liability and damages were denied by the district court judge and they
appealed the decision. In January 2007, the appellate court affirmed
the district court's decision on liability but remanded the case for a
more accurate calculation of lost pay. In its opinion, the $7${th}
Circuit reiterated its recommendation that in cases where the number of
eligible members of the protected group (White-males) exceeds the
number of available positions, the lost chance doctrine, which
originated in tort law, should be used, stating ``Loss of a chance is
illustrated by cases in which, as a result of a physician's negligent
failure to make a correct diagnosis, his patient's cancer is not
arrested, and he dies---but he probably would have died anyway. The
trier of fact will estimate the probability that the patient would have
survived but for the physician's negligence---say it is 25$\%$---and
will award that percentage of the damages the patient would have
received had it been certain that he would have survived but for the
negligence.'' [\textit{Alexander v. Milwaukee }, 474 F. 3d 437, 7th
Cir. (2007).]

The plaintiffs filed the case on June 27th, 2003, however, the
department would have known about the charge earlier. Although Chief
Jones remained in the position until 11/18/2003, often employers change
their employment practices after a charge has been formally filed
[\citet{FreGas00}]. Therefore, the court decided to rely
on data between 11/18/1996 and 5/31/2003 in both its liability
determination and compensation calculations, which is also used in our analysis.

Statistical tests for potential discrimination in sequential employment
decisions, for example, hiring, promotion or termination, are discussed
by \citet{Gas84}, \citet{Kad90}, \citet{GasGre95} and
Finkelstein and Levin (\citeyear{FinLev01}, pages~245--249). In \textit{Alexander v.
Milwaukee}, officers at the Sergeant rank for at least a year were
eligible to be promoted to lieutenant when an opening on the lieutenant
rank became available. The average length of time for a newly hired
police officer to become a lieutenant in our data is 18.21 years in
White-males and 15.76 in non-White-males ($p$-value${}=0.4058$). No claims of
discrimination were filed for promotion to lieutenant and this issue
was not mentioned in the legal decisions. Had there been evidence of
discrimination at lower ranks, the plaintiffs' lawyer would probably
have expanded the class of plaintiffs in the case. Therefore, it is
doubtful that the discriminatory practices affected that position.

First, both the promotion process and the retirement process are
modeled through Cox models. Seniority in the promotion process is
measured by the number of years the subject has served as a lieutenant.
The functional form for this time-varying covariate is quadratic. The
number of years since becoming eligible for retirement, which is also
time-varying, is used in the retirement model. Both these two
time-varying covariates are essentially the follow-up times in the two
processes, which usually cannot be used as covariates. However, because
our time axis is calendar time, people enter the processes on different
calendar dates and candidates in the risk set have different follow-up
times on the same calendar date. Therefore, we are able to estimate the
effects of number of years since lieutenant and number of years since
becoming eligible for retirement. Three time-invariant covariates are
also considered: membership in the protected group (White-male or not),
position (detective vs. police), number of years served in the police
force before becoming a lieutenant. While in some cases measures of
performance and disciplinary issues might have been considered, neither
the district nor the appellate court opinions mentioned any analysis
incorporating these factors. The defendant did not submit any data
about them, so it is unlikely they would differ much in the two groups.
The proportional hazards assumption between White-males and others is
examined by the parallel pattern of the log-cumulative-baseline-hazards
functions. The outputs of the two Cox models are given in Table \ref{tab3}. In
the model for the promotion risk among employees, the White-male factor
is highly significantly negative ($p$-value${}<0.001$), demonstrating that
White-male lieutenants had much lower promotion chances than
non-White-male employees with similar seniority and job assignment. The
length of time served as a lieutenant is also significant. The
coefficient for its square term is negative, which indicates that the
promotion chance among lieutenants initially increased with years of
service, but then reached a peak and declined afterward. In the
retirement data, 64 of the 112 individuals became eligible for
retirement during the period. Of them, 45 retired and 19 remained on
the job as of May 31st, 2003. Only three non-White-male officers
retired in the period and the White-male factor is not significant in
the retirement process. Each additional year after reaching eligibility
increases the retirement hazard by $13\%$ ($p$-value${}=0.017$), holding the
protected group membership, job assignment and number of years before
lieutenant constant. Also, the number of years served before lieutenant
is negatively correlated with retirement (Hazards Ratio${}={}$0.90,
$p$-value${}=0.007$). There were two lieutenants on Leave of Absence (LOA), a
Black male hired on 7/24/1978 and a White female hired on 7/30/1979,
which were treated as censored. The results in Table \ref{tab3} are robust when
the two LOA cases were deleted or treated as terminations.

%
\begin{table}
\caption{Analysis of \textit{Alexander v. Milwaukee}: Estimated
regression parameters from proportional hazards models. The columns
$e^{\widehat{\beta}_k}$ and $e^{\widehat{\theta}_k}$ are the hazards ratios}\label{tab3}
\begin{tabular*}{\textwidth}{@{\extracolsep{\fill}}lcd{2.2}ccd{2.2}cc@{}}
\hline
\multicolumn{1}{@{}l}{\textbf{Covariate,} $\bolds{Z_{ik}=X_{ik}}$} & &
\multicolumn{1}{c}{$\bolds{\widehat{\beta}_k}$}& \multicolumn{1}{c}{$\bolds{e^{\widehat{\beta
}_k}}$}& \multicolumn{1}{c}{$\bolds{p\mbox{-}\mathit{value}}$}&
\multicolumn{1}{c}{$\bolds{\widehat{\theta}_k}$} &\multicolumn{1}{c}{$\bolds{e^{\widehat{\theta}_k}}$} & \multicolumn{1}{c@{}}{$\bolds{p\mbox{-}\mathit{value}}$}
\\
\hline
White-male & &-2.13 &0.12 & $<$0.001 &0.18 &1.20&0.767\\
Detective & &-0.17 &0.84 & \phantom{$>$}0.611 &-0.23 &0.79&0.425\\
Years before lieutenant & &-0.01 &0.99 & \phantom{$>$}0.856 &-0.10 &0.90&0.007\\
Years since lieutenant & & 0.41 &1.51 & \phantom{$>$}0.012 &\multicolumn{1}{c}{--} &-- & --\\
Years since lieutenant$^2$ & &-0.02 &0.98 & \phantom{$>$}0.089 &\multicolumn{1}{c}{--} &-- & --\\
Years eligible for retire & &\multicolumn{1}{c}{--} &-- & \phantom{$>$}-- &0.12 &1.13&0.017\\
\hline
\end{tabular*}
\end{table}

\begin{figure}

\includegraphics{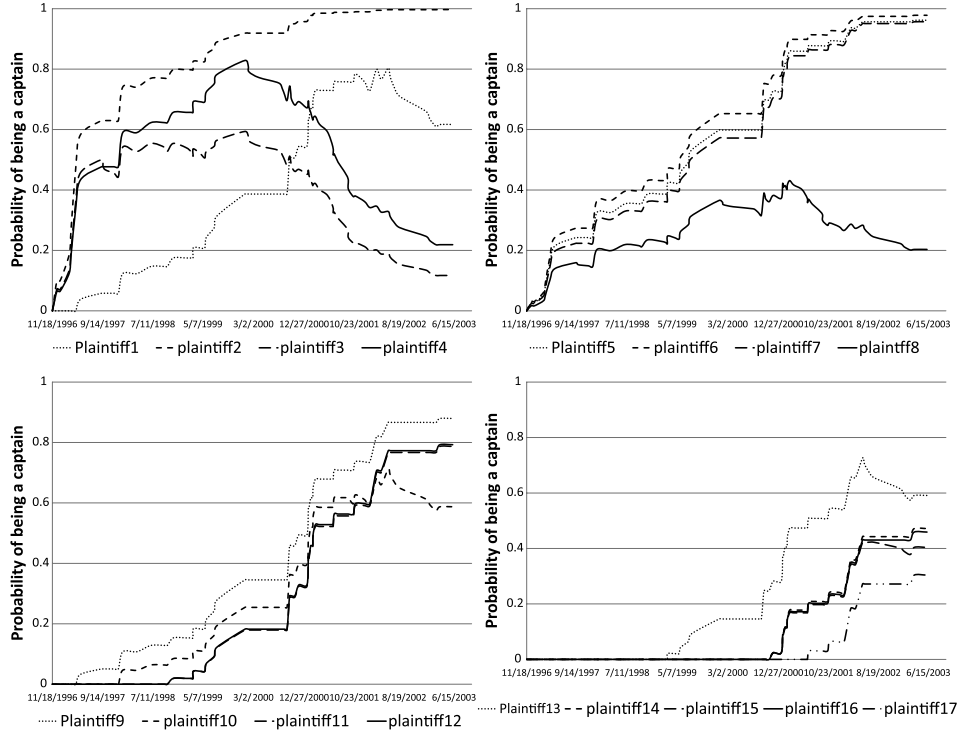}

\caption{Probability of being a captain and not retired,
$\operatorname{Pr}(P_{2i}^*\leq t, R_{2i}^*>t)=\break \widehat{S}^R(t|\tilde{X}_i(t),\tilde{Z}_i(t))
\times\{1-\widehat{S}^P(t|\tilde{X}_i(t),\tilde{Z}_i(t))\}$, for the 17
plaintiffs in the period [11/18/1996, 5/31/2003].}\label{fig1}
\end{figure}

The compensation estimates are based on the hypothetical scenario of no
discrimination, where the plaintiffs would have been treated the same
as non-White-males. The distributions of the three covariates in the
White-male and non-White-male groups overlap. The ranges of the numbers
of years before and after lieutenant in the plaintiffs are (10.97,
26.83) and (1.93, 12.88). The corresponding ranges in the
non-White-male group are (8.90, 26.60) and (0.21, 8.90). Therefore,
there were non-White-male members with similar qualifications as the
plaintiffs. The estimated probabilities of being a captain and not
retired, $\widehat{S}^R(t|\tilde{X}_i(t),\tilde{Z}_i(t))\{1-\widehat
{S}^P(t|\tilde{X}_i(t),\tilde{Z}_i(t))\}$, for the 17 plaintiffs are
plotted in Figure \ref{fig1}. The probabilities of being a captain and not
retired peak at around 5 years after becoming a lieutenant. The
promotion probabilities for plaintiffs 14, 15, 16 and 17 who were
lieutenants for less than two years never exceeded 0.5 during the
period. In contrast, plaintiffs who became lieutenants in 1996 and 1997
(e.g., plaintiffs 2, 5, 6, 7 and 9) have probabilities over 0.8 by May
2003. The probabilities for plaintiffs 3 and 4 peak around 2000 and
then decrease because their retirement probabilities increase after 2000.

%

%

Then, we estimate the expected lengths being a lieutenant, being a
captain and being retired restricted to the period [11/18/1996,
5/31/2003] for each plaintiff in Table \ref{tab4}. Consistent with the finding
of discrimination, the expected length of time one would remain a
lieutenant $\widehat{E}(T^{\mathrm{lt}})$ is always smaller than the observed
length~$T^{\mathrm{lt}}$. The expected number of months being a captain are
nonzero for every plaintiff, as they all lost some chance of being a
captain, although their observed months of being a captain during the
period are zero. Some plaintiffs have zero expected months of being
retired because they never became eligible for retirement during the
period. The sum of the expected months as a lieutenant, captain or in
retirement equals the sum of the corresponding observed months for each
plaintiff.\looseness=1

%
\begin{table}
\tabcolsep=0pt
\caption{Analysis of \textit{Alexander v. Milwaukee}: Columns $\widehat
{E}(T^{\mathrm{lt}})$, $\widehat{E}(T^{\mathrm{cap}})$ and $\widehat{E}(T^{\mathrm{rt}})$ are the
expected number of months being a lieutenant, a captain and in
retirement for each of the 17 plaintiffs under the hypothetical
nondiscriminatory scenario. Columns $T^{\mathrm{lt}}$, $T^{\mathrm{cap}}$ and $T^{\mathrm{rt}}$
are the corresponding observed number of months under the real life
scenario where the plaintiffs suffered discrimination}\label{tab4}
\begin{tabular*}{\textwidth}{@{\extracolsep{\fill}}lcccd{2.2}d{2.2}cd{2.2}d{2.2}d{2.2}@{}}
\hline
\multicolumn{1}{@{}l}{\textbf{Plaintiff}} &
\multicolumn{1}{c}{$\bolds{\widehat{E}(T^{\mathrm{lt}})}$}&
\multicolumn{1}{c}{$\bolds{\operatorname{SE}(\widehat{E}(T^{\mathrm{lt}}))}$} &
\multicolumn{1}{c}{$\bolds{T^{\mathrm{lt}}}$}&
\multicolumn{1}{c}{$\bolds{\widehat{E}(T^{\mathrm{cap}})}$}&
\multicolumn{1}{c}{$\bolds{\operatorname{SE}(\widehat{E}(T^{\mathrm{cap}}))}$} &
\multicolumn{1}{c}{$\bolds{T^{\mathrm{cap}}}$}&
\multicolumn{1}{c}{$\bolds{\widehat{E}(T^{\mathrm{rt}})}$} &
\multicolumn{1}{c}{$\bolds{\operatorname{SE}(\widehat{E}(T^{\mathrm{rt}}))}$}&
\multicolumn{1}{c}{$\bolds{T^{\mathrm{rt}}}$}
\\
\hline
\phantom{0}1 & 39.16& 5.98& 62.99& 30.84& 6.35& 0.00& 3.42& 2.10& 10.36 \\
\phantom{0}2 & 13.48& 4.80& 78.41& 64.96& 4.80& 0.00& 0.00& 0.00& 0.00 \\
\phantom{0}3 & 15.52& 5.85& 78.41& 30.44&12.59& 0.00&32.48&10.26& 0.00 \\
\phantom{0}4 & 19.17& 6.71& 78.41& 40.47&10.52& 0.00&18.81&06.35& 0.00 \\
\phantom{0}5 & 32.68& 6.38& 78.41& 45.76& 6.38& 0.00& 0.00& 0.00& 0.00 \\
\phantom{0}6 & 29.85& 5.03& 78.41& 48.59& 5.03& 0.00& 0.00& 0.00& 0.00 \\
\phantom{0}7 & 34.03& 6.51& 78.41& 44.42& 6.51& 0.00& 0.00& 0.00& 0.00 \\
\phantom{0}8 & 31.13& 6.87& 78.41& 19.86& 9.86& 0.00&26.56&10.22& 0.00 \\
\phantom{0}9 & 41.39& 5.36& 72.92& 31.56& 5.36& 0.00& 0.00& 0.00& 0.00 \\
10 & 39.45& 5.52& 66.25& 23.61& 5.98& 0.00& 3.22& 1.84& 0.00 \\
11 & 32.42& 4.08& 54.74& 22.36& 4.08& 0.00& 0.00& 0.00& 0.00 \\
12 & 18.90& 1.68& 23.21& 4.34& 1.68& 0.00& 0.00& 0.00& 0.00 \\
13 & 31.04& 4.83& 51.75& 19.13& 4.96& 0.00& 1.61& 1.02& 0.00 \\
14 & 20.91& 2.50& 30.12& 9.24& 2.50& 0.00& 0.00& 0.00& 0.00 \\
15 & 21.11& 2.99& 30.12& 8.58& 3.02& 0.00& 0.49& 0.33& 0.00 \\
16 & 21.17& 2.60& 30.12& 8.98& 2.60& 0.00& 0.00& 0.00& 0.00 \\
17 & 32.19& 3.78& 54.74& 22.59& 3.78& 0.00& 0.00& 0.00& 0.00 \\
\hline
\end{tabular*}
\end{table}

The compensatory damages proposed by the District Court were calculated
according to a set of specific instructions. For each plaintiff the
jury selected one date for the possible promotion of each plaintiff
along with an estimated probability that the plaintiff would have been
promoted that day. Contrary to the time-varying probability estimates
in Figure \ref{fig1} which incorporate relevant covariates, the probabilities
used in the District Court were either 0.50 (plaintiffs 1 and 10) or 0.80
(all other plaintiffs). The retirement dates for each plaintiff used in
the jury's calculation were the observed ones (if the plaintiff retired
during the period) or the first day he became eligible (if the
plaintiff did not retire during the period). Table~\ref{tab5} compares the
estimated numbers of months that plaintiffs 1 to 13 would have served
in different levels (Lieutenant, Captain and retirement) obtained from
the proposed method and those determined by the original jury. The
jury's estimates for Plaintiffs 13 to 17 were not available to us.
Although the estimates are close in a few cases (plaintiffs 1 and 10),
there are a number of substantial discrepancies (plaintiffs 2, 6, 7, 8,
9) and noticeable ones (plaintiffs 3, 4, 5, 11, 12, 13).

%
\begin{table}
\tabcolsep=0pt
\caption{Analysis of \textit{Alexander v. Milwaukee}: Comparison of the
proposed job and retirement duration estimates and the previous
estimates reversed by the circuit court (months)}\label{tab5}
\begin{tabular*}{\textwidth}{@{\extracolsep{\fill}}lcccd{2.2}d{2.2}cd{2.2}d{2.2}@{}}
\hline
& \multicolumn{2}{c}{\textbf{Lieutenant}}&&\multicolumn
{2}{c}{\textbf{Captain}}&&\multicolumn{2}{c}{\textbf{Retired}}\\[-6pt]
& \multicolumn{2}{c}{\hrulefill}&&\multicolumn
{2}{c}{\hrulefill}&&\multicolumn{2}{c@{}}{\hrulefill}\\
\textbf{Plaintiff} &\multicolumn{1}{c}{$\bolds{\widehat{E}(T_{\mathrm{proposed}})}$} &\multicolumn{1}{c}{$\bolds{\widehat{E}(T_{\mathrm{jury}}})$}
&&\multicolumn{1}{c}{$\bolds{\widehat{E}(T_{\mathrm{proposed}})}$} &\multicolumn{1}{c}{$\bolds{\widehat{E}(T_{\mathrm{jury}})}$} &&\multicolumn{1}{c}{$\bolds{\widehat
{E}(T_{\mathrm{proposed}})}$} &\multicolumn{1}{c@{}}{$\bolds{\widehat{E}(T_{\mathrm{jury}})}$}
\\\hline
\phantom{0}1 & 39.16& 30.64&& 30.84& 32.38&& 3.42&10.39\\
\phantom{0}2 & 13.48& 52.83&& 64.96& 25.58&& 0.00& 0.00\\
\phantom{0}3 & 15.52& 39.78&& 30.44& 38.63&&32.48& 0.00\\
\phantom{0}4 & 19.17& 31.13&& 40.47& 47.28&&18.81& 0.00\\
\phantom{0}5 & 32.68& 26.70&& 45.76& 51.72&& 0.00& 0.00\\
\phantom{0}6 & 29.85& 70.95&& 48.59& 7.46&& 0.00& 0.00\\
\phantom{0}7 & 34.03& 52.83&& 44.42& 25.58&& 0.00& 0.00\\
\phantom{0}8 & 31.13& 44.98&& 19.86& 32.55&&26.56& 0.00\\
\phantom{0}9 & 41.39& 51.42&& 31.56& 21.50&& 0.00& 0.00\\
10 & 39.45& 43.63&& 23.61& 22.62&& 3.22& 0.00\\
11 & 32.42& 36.16&& 22.36& 18.58&& 0.00& 0.00\\
12 & 18.90& 15.75&& 4.34& 7.46&& 0.00& 0.00\\
13 & 31.04& 41.65&& 19.13& 10.09&& 1.61& 0.00\\
\hline
\end{tabular*}
\end{table}

\section{Discussion}\label{sec6}

This paper provides a method for estimating the length of time
plaintiffs who were discriminated against would have been in a higher
position. In the motivating example the promotion process was of
primary interest and retirement terminates an employee's eligibility
for further promotion. Estimators of the mean durations for remaining
at the lower position, being at the higher position and being retired,
restricted to the relevant time interval, are obtained. The asymptotic
distributions of the restricted mean life time estimators are derived
and shown to perform well in finite samples in our simulation studies.
The proposed compensation estimators are obtained by assuming a
non-White-male counterpart with similar qualifications for each
plaintiff. The methodology is applied to obtain the three restricted
mean job durations absent discrimination and their standard errors for
each plaintiff in \textit{Alexander v. Milwaukee}.

The context of legal cases is different from the causal inference
widely used in epidemiologic studies and clinical trials [\citet{Rub74};
\citet{HavNag05}]. First, our ultimate goal is not to derive
the causal relationship between the White-male factor and promotion
risks. As emphasized by Judge Easterbrook in \textit{Biondo v. City of
Chicago} (382 F.3d 680, 7th Cir. 2004), the purpose of awarding damages
is to put the plaintiffs in the position they should have been in
during the period of time they deserve compensation. The ``gold
standard'' for determining the compensation due a White-male lieutenant
who suffered discrimination in promotion is based on the promotion and
retirement probabilities of a similarly qualified non-White-male
lieutenant during the period of discrimination. Second, the standard of
proof in civil cases is the preponderance of the evidence or ``more
likely than not.'' It is less stringent than the criteria scientific
research uses to determine a causal relationship. Third, only
characteristics which were actually considered as promotion criteria
(e.g., seniority, performance, education, exam scores{\ldots}) will be
considered as potential confounders. For a hypothetical example,
suppose there had been another covariate---accent in speaking English---that was
significantly correlated with the White-male factor and the
promotion risks. However, as long as accent was not used as a criterion
in the promotion process, we should not adjust for it as a confounder.
Fourth, if there are other potential confounders that were used as
promotion criteria, presumably the defendant would include them in
their analysis. Indeed, once a plaintiff submits a reasonable
statistical analysis incorporating the main covariates, the defendant
cannot simply suggest another omitted variable that might explain the
disparity. Rather the defendant should incorporate subject-level
information on that factor into their analysis. Last, in contrast to
areas employing causal inference (e.g., clinical trials), where one
wishes to generalize the results to a much larger population, the
disparity estimated between White-male and non-White-males in the
Milwaukee Police Department in the period under study is not
generalizable to other departments or other time periods.

When the wages and pension benefits are stable over time, the
compensatory damages can be calculated as $\mathrm{Wage}_{\mathrm{lt}}\widehat
{E}(T^{\mathrm{lt}})+\mathrm{Wage}_{\mathrm{cap}}\widehat{E}(T^{\mathrm{cap}})+\mathrm{Pension}\widehat{E}(T^{\mathrm{rt}})$
minus the plaintiffs' actual earnings. However, wages and pensions
often change over time. Then one can calculate the weighted average of
the lieutenant wage, the captain wage and the retirement pension, at
each time point, where the weights are the probabilities of being a
lieutenant, a captain and retired. These weighted averages are
integrated over time, that is, $\int_{\tau_0}^{\tau_1}\{
\mathrm{Wage}_{\mathrm{lt}}\operatorname{Pr}(\mathrm{lt})+\mathrm{Wage}_{\mathrm{cap}}\operatorname{Pr}(\mathrm{cap})+
\mathrm{Pension}\operatorname{Pr}(\mathrm{rt})\}\,dt$. This type of
compensation calculation incorporating salary information is described
in Pan and Gastwirth (\citeyear{PanGas}), in a \textit{simpler} context where only
point estimates are given.

The time period over which the economic damages will be paid depends on
the specific facts of each case because compensation ends when the
effect of the discriminatory practices ceases and is determined by the
court. If seniority has a major role as in \textit{Alexander v.
Milwaukee}, the time when the discrimination effect on a particular
plaintiff ends may depend on the promotions and retirements of more
senior plaintiffs. Thus, for the purpose of compensation, it is
desirable for courts to require employers to provide pay data beyond
the period of discrimination used in the liability stage.

Although the problem addressed here arose in the context of a legal
case, predicting the durations before and after the event of interest
in the presence of a terminating event occurs in other applications.
For example, when banks merge, the value of the bank being taken over
depends on the expected durations of the existing accounts. Each
account may remain at the same level, be upgraded to a higher type or
be closed. The latter two correspond to the promotion and retirement in
our motivating example. In the academic job market, people are often
interested in the length of time individuals spent as a postdoc before
obtaining a regular position. Some postdocs eventually opt to take a
job outside the subject of their doctoral degree, which is the
terminating event.

\begin{appendix}\label{app}

\section*{Appendix}
\subsection{Plot of hypothetical scenarios of the promotion and
retirement processes}

\begin{figure}

\includegraphics{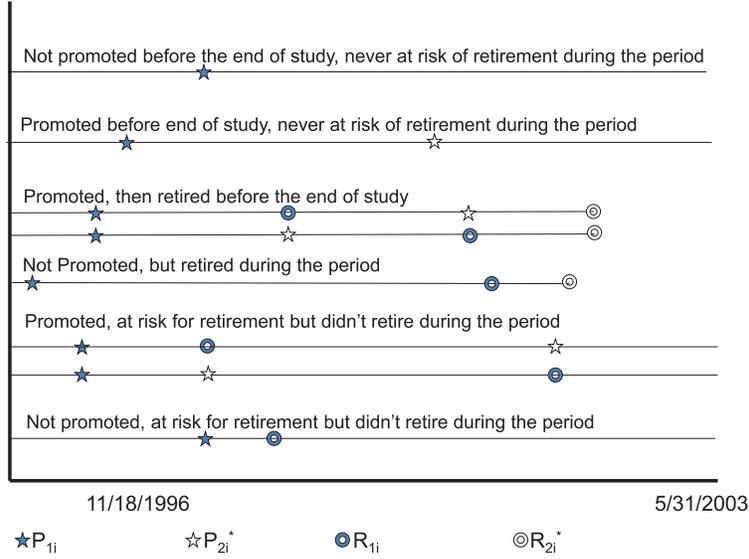}

\caption{Possible scenarios for the promotion and retirement
processes with hypothetical $P_{1i}, P_{2i}^*, R_{1i}, R_{2i}^*$ values.}\label{fig2}
\end{figure}

For the plot of hypothetical scenarios for the promotion and
retirement processes, see Figure \ref{fig2}.

%

\subsection{Proofs of the theorems}

Under conditions (a) to (f), the almost sure convergence of $\widehat
{\beta}$ to $\beta_0$ and $\widehat{\theta}$ to $\theta_0$ holds from
the Empirical Central Limit theorem [\citet{Pol90}]. Furthermore,
$\widehat{\Lambda}_0^P(t;\beta_0)\stackrel{\mathrm{a.s.}}{\longrightarrow}\Lambda
_0^P(t)$ and $\widehat{\Lambda}_0^R(t;\theta_0)\stackrel
{\mathrm{a.s.}}{\longrightarrow}\Lambda_0^R(t)$ for all $t\in[\tau_0,\tau_1]$
[\citet{AndGil82}]. By the continuous mapping theorem [\citet{SenSin93}],
\[
\int_{\tau_0}^{\tau_1} \bigl\{1-\widehat{S}^P
\bigl(t|\tilde{X}_i(t) \bigr) \bigr\}\widehat{S}^R
\bigl(t| \tilde{Z}_i(t) \bigr)\,dt\stackrel{\mathrm{a.s.}} {
\longrightarrow}E \bigl(T^{\mathrm{cap}}|\tilde{X}_i(t),
\tilde{Z}_i(t) \bigr).
\]

To derive the variance, $n^{{1}/{2}}\{\widehat{S}^P(t|\tilde
{X}_i(t))-S^P(t|\tilde{X}_i(t))\}$ is decomposed into two parts:
%
%
\begin{eqnarray}
\label{2}
&&
n^{{1}/{2}} \bigl\{\widehat{S}^P \bigl(t|
\tilde{X}_i(t) \bigr)-S^P \bigl(t|\tilde{X}_i(t)
\bigr) \bigr\} \nonumber\\
&&\qquad=n^{{1}/{2}} \bigl\{\widehat{S}^P \bigl(t;
\widehat{ \beta}, d\widehat{\Lambda}_0^P(t)|
\tilde{X}_i(t) \bigr)-\widehat{S}^P \bigl(t;
\beta_0, d\widehat{\Lambda }_0^P(t)|
\tilde{X}_i(t) \bigr) \bigr\}
\\
&&\qquad\quad{}+n^{{1}/{2}} \bigl\{\widehat{S}^P \bigl(t;
\beta_0, d\widehat{\Lambda}_0^P(t)|
\tilde{X}_i(t) \bigr)-S^P \bigl(t|\tilde{X}_i(t)
\bigr) \bigr\}.\nonumber
\end{eqnarray}

Apply a Taylor expansion to the first term on the right side of (\ref
{2}) around
$\beta_0$. As $n\rightarrow\infty$,
%
%
\begin{eqnarray}
\label{2.1deri}
 &&n^{{1}/{2}} \bigl\{\widehat{S}^P \bigl(t;
\widehat{\beta}, d\widehat{\Lambda}_0^P(t)|
\tilde{X}_i(t) \bigr)-\widehat{S}^P \bigl(t;
\beta_0, d\widehat{\Lambda}_0^P(t)|
\tilde{X}_i(t) \bigr) \bigr\}
\nonumber
\\[2pt]
&&\qquad=\frac{\partial{\widehat{S}^P}(t;\beta,d\widehat{\Lambda
}_0^P(t)|\tilde{X}_i(t))}{\partial\beta'}\Big|_{\beta=\beta_*}n^{
{1}/{2}}(\widehat{\beta}-
\beta_0)
\\
&&\qquad=-\widehat{S}^P \bigl(t;\beta_*,d\widehat{\Lambda}_0^P(t)|
\tilde{X}_i(t) \bigr)\nonumber\\
&&\qquad\quad{}\times\int_{P_{1i}}^t
\bigl[e^{\beta'_*\tilde{X}_i(u)} \bigl\{\tilde{X}_i(u)-\overline {X}(u;\beta_*)
\bigr\}\,d\widehat{\Lambda}^P_0(u;\beta_*)
\bigr]n^{{1}/{2}}(\widehat{\beta}-\beta_0),\nonumber
\end{eqnarray}
where $\beta_{*}$ lies between $\widehat{\beta}$ and $\beta_{0}$.
Furthermore, another Taylor expansion of the score function
$U^P(\widehat{\beta})$ around $\beta_0$ yields
%
%
\begin{equation}
\label{beta} n^{{1}/{2}}(\widehat{\beta}-\beta_{0})= \bigl
\{A^P(\beta_0) \bigr\} ^{-1}n^{-{1}/{2}}\sum
_{i=1}^nU_i^P(
\beta_{0}),
\end{equation}
where $U_i^P(\beta)=\int_{\tau_0}^{\tau_1}\{X_i(t)-\overline{X}(t;\beta
)\} \,dN_i^P(t)$. Combining (\ref{2.1deri}) and (\ref{beta}) yields
%
%
\begin{eqnarray}
\label{key21}&& n^{{1}/{2}} \bigl\{\widehat{S}^P \bigl(t;
\widehat{ \beta},d\widehat{\Lambda}_0^P(t)|
\tilde{X}_i(t) \bigr)-\widehat{S}^P \bigl(t;
\beta_0,d\widehat{\Lambda }_0^P(t)|
\tilde{X}_i(t) \bigr) \bigr\}
\nonumber
\\[-8pt]
\\[-8pt]
\nonumber
&&\qquad=n^{-{1}/{2}}\sum
_{i=1}^n \xi^P_{i1}(t),
\end{eqnarray}
where $\xi^P_{i1}(t)=-\widehat{S}^P(t|\tilde{X}_i(t))\int_{P_{1i}}^t[e^{\beta_0'\tilde{X}_i(u)}\{\tilde{X}_i(u)-\overline
{X}(u;\beta_0)\}\,d\widehat{\Lambda}^P_0(u;\beta_0)]\times\break \{A^P(\beta_0)\}
^{-1}U_i^P(\beta_0)$.

For the second term on the right side of (\ref{2}), when $n\rightarrow
\infty$,
\begin{eqnarray*}
&&n^{{1}/{2}} \bigl\{\widehat{S}^P \bigl(t;
\beta_0,d \widehat{\Lambda}_0^P(t)|
\tilde{X}_i(t) \bigr)-S^P \bigl(t|\tilde{X}_i(t)
\bigr) \bigr\}
\\
&&\qquad=n^{{1}/{2}} \biggl[\exp \biggl\{-\int_{P_{1i}}^td
\widehat{\Lambda}^P_0(u;\beta_0)e^{\beta'_0\tilde{X}_{i}(u)}
\biggr\}-\exp \biggl\{-\int_{P_{1i}}^td\Lambda
_0^P(u)e^{\beta'_0\tilde{X}_{i}(u)} \biggr\} \biggr]
\\
&&\qquad=-S^P \bigl(t|\tilde{X}_i(t) \bigr)n^{{1}/{2}}
\int_{P_{1i}}^te^{\beta'_0\tilde
{X}_{i}(u)} \bigl\{d\widehat{
\Lambda}_0^P(u;\beta_0)-d
\Lambda_0^P(u) \bigr\},
\end{eqnarray*}
while
\begin{eqnarray*}
&&n^{{1}/{2}} \bigl\{d\widehat{\Lambda}_0^P(u;
\beta_0)-d\Lambda_0^P(u) \bigr\}\\
&&\qquad=n^{-{1}/{2}} \biggl\{\frac{\sum_{i=1}^n\,
dN_{i}^P(u)}{S^{(0)}_P(u;\beta_0)}-d\Lambda_0^P(u)
\biggr\}
\\
&&\qquad=n^{-{1}/{2}}\sum_{i=1}^n
\frac{dM_{i}^P(u;\beta
_0)}{S^{(0)}_P(u;\beta_0)}
\\
&&\qquad=n^{-{1}/{2}}\sum_{i=1}^n
\frac{dM_{i}^P(u;\beta
_0)}{s^{(0)}_P(u;\beta_0)}
\\
&&\qquad\quad{}+n^{-{1}/{2}}\sum_{i=1}^n
\bigl[S^{(0)}_P(u;\beta_0)^{-1}-s^{(0)}_P(u;
\beta_0)^{-1} \bigr]\,dM_{i}^P(u;
\beta_0),
\end{eqnarray*}
where $s^{(0)}_P(u;\beta)=\lim_{n\rightarrow\infty}S^{(0)}_P(u;\beta)$
and $dM_i^P(u;\beta)=dN_i^P(u)-Y_i^P(u)\times e^{\beta'X_i(u)}\,d\Lambda_0^P(u)$.
The second term
\[
n^{-{1}/{2}}\sum_{i=1}^n\bigl[S^{(0)}_P(u;\beta
_0)^{-1}-s^{(0)}_P(u;\beta_0)^{-1}\bigr]\,dM_{i}^P(u;\beta_0)
\]
converges to zero,
by the strong convergence of $S^{(0)}(r;\beta_0)$
to $s^{(0)}(r;\beta_0)$, the continuous mapping theorem and the
uniform strong law of large numbers. Therefore, as $n\rightarrow\infty
$,
\begin{eqnarray*}
\label{key22} n^{{1}/{2}} \bigl\{d\widehat{\Lambda}^P(t;
\beta_0)-d\Lambda(t)^P \bigr\}&\approx&n^{-{1}/{2}}
\sum_{i=1}^n\frac{dM_{i}^P(t;\beta
_0)}{s^{(0)}_P(t;\beta_0)},
\\
n^{{1}/{2}} \bigl\{\widehat{S}^P \bigl(t;\beta_0,d
\widehat{\Lambda}_0^P(t)|\tilde{X}_i(t)
\bigr)-S^P \bigl(t|\tilde{X}_i(t) \bigr) \bigr
\}&=&n^{-{1}/{2}}\sum_{i=1}^n
\xi^P_{i2}(t),
\end{eqnarray*}
where
\[
\xi_{i2}^P(t)=-S^P \bigl(t|
\tilde{X}_i(t) \bigr)\int_{P_{1i}}^te^{\beta'_0\tilde{X}_{i}(u)}
\frac{dM_{i}^P(u;\beta
_0)}{s^{(0)}_P(u;\beta_0)}.
\]

In summary, combining (\ref{key21}) and (\ref{key22}), when
$n\rightarrow\infty$,
%
%
\begin{equation}
\label{xi1} n^{{1}/{2}} \bigl\{\widehat{S}^P \bigl(t|
\tilde{X}_i(t) \bigr)-S^P \bigl(t|\tilde{X}_i(t)
\bigr) \bigr\} =n^{-{1}/{2}}\sum_{i=1}^n
\bigl\{\xi_{i1}^P(t)+\xi_{i2}^P(t)
\bigr\}.
\end{equation}

Similarly,
%
%
\begin{equation}
\label{xi2} n^{{1}/{2}} \bigl\{\widehat{S}^R \bigl(t|
\tilde{Z}_i(t) \bigr)-S^R \bigl(t|\tilde{Z}_i(t)
\bigr) \bigr\} =n^{-{1}/{2}}\sum_{i=1}^n
\bigl\{\xi_{i1}^R(t)+\xi_{i2}^R(t)
\bigr\},
\end{equation}
where
\begin{eqnarray*}
\xi^R_{i1}(t)&=&-\widehat{S}^R \bigl(t|
\tilde{Z}_i(t) \bigr)\\
&&{}\times\int_{R_{1i}}^t
\bigl[e^{\theta_0'\tilde{Z}_i(u)} \bigl\{\tilde{Z}_i(u)-\overline{Z}(u;
\theta_0) \bigr\}\,d\widehat{\Lambda}^R_0(u;
\theta_0) \bigr]{A^R(\theta_0)}^{-1}U_i^R(
\theta_0)
\\
\xi_{i2}^R(t)&=&-S^R \bigl(t|
\tilde{Z}_i(t) \bigr)\int_{R_{1i}}^te^{\theta'_0\tilde
{Z}_{i}(u)}
\frac{dM_{i}^R(u;\theta_0)}{s^{(0)}_R(u;\theta_0)}.
\end{eqnarray*}
Combining (\ref{xi1}) and (\ref{xi2}), we get
\begin{eqnarray*}
&&n^{{1}/{2}} \bigl\{ \bigl(1-\widehat{S}^P \bigl(t|
\tilde{X}_i(t) \bigr) \bigr)\widehat{S}^R \bigl(t|
\tilde{Z}_i(t) \bigr)- \bigl(1-S^P \bigl(t|
\tilde{X}_i(t) \bigr) \bigr)S^R \bigl(t|
\tilde{Z}_i(t) \bigr) \bigr\}
\\
&&\qquad=n^{{1}/{2}} \bigl\{ \bigl(1-\widehat{S}^P \bigl(t|
\tilde{X}_i(t) \bigr) \bigr)\widehat{S}^R \bigl(t|
\tilde{Z}_i(t) \bigr)- \bigl(1-S^P \bigl(t|
\tilde{X}_i(t) \bigr) \bigr)\widehat{S}^R \bigl(t|\tilde
{Z}_i(t) \bigr) \bigr\}
\\
&&\qquad\quad{}+n^{{1}/{2}} \bigl\{ \bigl(1-S^P \bigl(t|
\tilde{X}_i(t) \bigr) \bigr)\widehat{S}^R \bigl(t|
\tilde{Z}_i(t) \bigr)- \bigl(1-S^P \bigl(t|
\tilde{X}_i(t) \bigr) \bigr)S^R \bigl(t|
\tilde{Z}_i(t) \bigr) \bigr\}
\\
&&\qquad=n^{-{1}/{2}}\sum_{i=1}^n \bigl[
\widehat{S}^R \bigl\{t|\tilde{Z}_i(t) \bigr\} \bigl\{ -
\xi_{i1}^P(t)-\xi_{i2}^P(t) \bigr\}\\
&&\qquad\quad\qquad\hspace*{20pt}{}+
\bigl\{1-S^P \bigl(t|\tilde{X}_i(t) \bigr) \bigr\} \bigl
\{ \xi_{i1}^R(t)+\xi_{i2}^R(t) \bigr\}
\bigr].
\end{eqnarray*}
Integrating over $[\tau_0,\tau_1]$,
\[
n^{{1}/{2}} \bigl\{\widehat{E} \bigl(T^{\mathrm{cap}}|\tilde{X}_i(t),
\tilde{Z}_i(t) \bigr)-E \bigl(T^{\mathrm{cap}}|\tilde{X}_i(t),
\tilde{Z}_i(t) \bigr) \bigr\}=n^{-
{1}/{2}}\sum
_{i=1}^n\int_{\tau_0}^{\tau_1}
\xi_i(t)\,dt,
\]
where
\[
\xi_i(t)=\bigl\{\widehat{S}^R \bigl(t|
\tilde{Z}_i(t) \bigr) \bigl(-\xi_{i1}^P(t)-\xi
_{i2}^P(t) \bigr)+ \bigl(1-S^P \bigl(t|
\tilde{X}_i(t) \bigr) \bigr) \bigl(\xi_{i1}^R(t)+
\xi_{i2}^R(t) \bigr) \bigr\}.
\]
By the empirical process theory [\citet{Pol90}; Van der Vaart and
Wellner (\citeyear{vanWel96})], $n^{{1}/{2}}\{\widehat{E}(T^{\mathrm{cap}}|\tilde
{X}_i(t),\tilde{Z}_i(t))-E(T^{\mathrm{cap}}|\tilde{X}_i(t),\tilde{Z}_i(t))\}$
converges weakly to a mean-zero Gaussian process with variance $E[\{\int_{\tau_0}^{\tau_1}\xi_i(t)\,dt\}^2]$,
which can be estimated by replacing
parameters with their empirical estimates and expectations with sample
averages. The asymptotic distributions of $\widehat{E}(T^{\mathrm{lt}}|\tilde
{X}_i(t),\tilde{Z}_i(t))$ and $\widehat{E}(T^{\mathrm{rt}}|\tilde{X}_i(t),\tilde
{Z}_i(t))$ can be derived similarly.
\end{appendix}

%
\section*{Acknowledgments}
We appreciate the constructive comments and suggestions from the
referee and Associate Editor, which greatly improved our manuscript.


%

\printaddresses


\begin{thebibliography}{31}

\bibitem[\protect\citeauthoryear{Andersen and Gill}{1982}]{AndGil82}
%
\begin{barticle}[mr]
\bauthor{\bsnm{Andersen},~\bfnm{P.~K.}\binits{P.~K.}} \AND
\bauthor{\bsnm{Gill},~\bfnm{R.~D.}\binits{R.~D.}}
(\byear{1982}).
\btitle{Cox's regression model for counting processes: A large sample study}.
\bjournal{Ann. Statist.}
\bvolume{10}
\bpages{1100--1120}.
\bid{issn={0090-5364}, mr={0673646}}
\bptok{imsref}%
\end{barticle}
%
\endbibitem

\bibitem[\protect\citeauthoryear{Chen and Tsiatis}{2001}]{CheTsi01}
%
\begin{barticle}[mr]
\bauthor{\bsnm{Chen},~\bfnm{Pei-Yun}\binits{P.-Y.}} \AND
\bauthor{\bsnm{Tsiatis},~\bfnm{Anastasios~A.}\binits{A.~A.}}
(\byear{2001}).
\btitle{Causal inference on the difference of the restricted mean lifetime
between two groups}.
\bjournal{Biometrics}
\bvolume{57}
\bpages{1030--1038}.
\bid{doi={10.1111/j.0006-341X.2001.01030.x}, issn={0006-341X}, mr={1950418}}
\bptok{imsref}%
\end{barticle}
%
\endbibitem

\bibitem[\protect\citeauthoryear{Cook and Lawless}{1997}]{CooLaw97}
%
\begin{barticle}[auto:STB|2013/05/03|14:24:43]
\bauthor{\bsnm{Cook},~\bfnm{R.~J.}\binits{R.~J.}} \AND
\bauthor{\bsnm{Lawless},~\bfnm{J.~F.}\binits{J.~F.}}
(\byear{1997}).
\btitle{Marginal analysis of recurrent events and a terminating event}.
\bjournal{Stat. Med.}
\bvolume{16}
\bpages{911--924}.
\bptok{imsref}%
\end{barticle}
%
\endbibitem

\bibitem[\protect\citeauthoryear{Cox}{1972}]{Cox72}
%
\begin{barticle}[mr]
\bauthor{\bsnm{Cox},~\bfnm{D.~R.}\binits{D.~R.}}
(\byear{1972}).
\btitle{Regression models and life-tables}.
\bjournal{J. R. Stat. Soc. Ser. B Stat. Methodol.}
\bvolume{34}
\bpages{187--220}.
\bid{issn={0035-9246}, mr={0341758}}
\bptnote{check related}%
\bptok{imsref}%
\end{barticle}
%
\endbibitem

\bibitem[\protect\citeauthoryear{Cox}{1975}]{Cox75}
%
\begin{barticle}[mr]
\bauthor{\bsnm{Cox},~\bfnm{D.~R.}\binits{D.~R.}}
(\byear{1975}).
\btitle{Partial likelihood}.
\bjournal{Biometrika}
\bvolume{62}
\bpages{269--276}.
\bid{issn={0006-3444}, mr={0400509}}
\bptok{imsref}%
\end{barticle}
%
\endbibitem

\bibitem[\protect\citeauthoryear{Fine, Jiang and Chappell}{2001}]{FinJiaCha01}
%
\begin{barticle}[mr]
\bauthor{\bsnm{Fine},~\bfnm{J.~P.}\binits{J.~P.}},
\bauthor{\bsnm{Jiang},~\bfnm{H.}\binits{H.}} \AND
\bauthor{\bsnm{Chappell},~\bfnm{R.}\binits{R.}}
(\byear{2001}).
\btitle{On semi-competing risks data}.
\bjournal{Biometrika}
\bvolume{88}
\bpages{907--919}.
\bid{doi={10.1093/biomet/88.4.907}, issn={0006-3444}, mr={1872209}}
\bptok{imsref}%
\end{barticle}
%
\endbibitem

\bibitem[\protect\citeauthoryear{Finkelstein and Levin}{2001}]{FinLev01}
%
\begin{bbook}[auto:STB|2013/05/03|14:24:43]
\bauthor{\bsnm{Finkelstein},~\bfnm{M.~O.}\binits{M.~O.}} \AND
\bauthor{\bsnm{Levin},~\bfnm{B.}\binits{B.}}
(\byear{2001}).
\btitle{Statistics for Lawyers}.
\bpublisher{Springer}, \blocation{New York}.
\bptok{imsref}%
\end{bbook}
%
\endbibitem


\bibitem[\protect\citeauthoryear{Freidlin and Gastwirth}{2000}]{FreGas00}
%
\begin{barticle}[auto:STB|2013/05/03|14:24:43]
\bauthor{\bsnm{Freidlin},~\bfnm{B.}\binits{B.}} \AND
\bauthor{\bsnm{Gastwirth},~\bfnm{J.~L.}\binits{J.~L.}}
(\byear{2000}).
\btitle{Changepoint tests designed for the analysis of hiring data
arising in
employment discrimination cases}.
\bjournal{J. Bus. Econom. Statist.}
\bvolume{18}
\bpages{315--322}.
\bptok{imsref}%
\end{barticle}
%
\endbibitem

\bibitem[\protect\citeauthoryear{Gastwirth}{1984}]{Gas84}
%
\begin{barticle}[auto:STB|2013/05/03|14:24:43]
\bauthor{\bsnm{Gastwirth},~\bfnm{J.~L.}\binits{J.~L.}}
(\byear{1984}).
\btitle{Statistical methods for analyzing claims of employment decisions}.
\bjournal{Industrial and Labor Relations Review}
\bvolume{38}
\bpages{75--86}.
\bptok{imsref}%
\end{barticle}
%
\endbibitem

\bibitem[\protect\citeauthoryear{Gastwirth and Greenhouse}{1995}]{GasGre95}
%
\begin{barticle}[pbm]
\bauthor{\bsnm{Gastwirth},~\bfnm{J.~L.}\binits{J.~L.}} \AND
\bauthor{\bsnm{Greenhouse},~\bfnm{S.~W.}\binits{S.~W.}}
(\byear{1995}).
\btitle{Biostatistical concepts and methods in the legal setting}.
\bjournal{Stat. Med.}
\bvolume{14}
\bpages{1641--1653}.
\bid{issn={0277-6715}, pmid={7481200}}
\bptok{imsref}%
\end{barticle}
%
\endbibitem

\bibitem[\protect\citeauthoryear{Goldstein}{2011}]{Gol}
%
\begin{bmisc}[auto:STB|2013/05/03|14:24:43]
\bauthor{\bsnm{Goldstein},~\bfnm{J.}\binits{J.}}
(\byear{2011}).
\bhowpublished{Still on patrol after two decades, valued
but rare. \textit{The New York Times} \textbf{December 13} A30}.
\bptok{imsref}%
\end{bmisc}
%
\endbibitem

\bibitem[\protect\citeauthoryear{Haviland and Nagin}{2005}]{HavNag05}
%
\begin{barticle}[mr]
\bauthor{\bsnm{Haviland},~\bfnm{Amelia~M.}\binits{A.~M.}} \AND
\bauthor{\bsnm{Nagin},~\bfnm{Daniel~S.}\binits{D.~S.}}
(\byear{2005}).
\btitle{Causal inferences with group based trajectory models}.
\bjournal{Psychometrika}
\bvolume{70}
\bpages{557--578}.
\bid{doi={10.1007/s11336-004-1261-y}, issn={0033-3123}, mr={2272504}}
\bptok{imsref}%
\end{barticle}
%
\endbibitem

\bibitem[\protect\citeauthoryear{Kadane}{1990}]{Kad90}
%
\begin{barticle}[auto:STB|2013/05/03|14:24:43]
\bauthor{\bsnm{Kadane},~\bfnm{J.~B.}\binits{J.~B.}}
(\byear{1990}).
\btitle{A statistical analysis of adverse impact of employer decisions}.
\bjournal{J.~Amer. Statist. Assoc.}
\bvolume{85}
\bpages{925--933}.
\bptok{imsref}%
\end{barticle}
%
\endbibitem

\bibitem[\protect\citeauthoryear{Kadane and Woodworth}{2004}]{KadWoo04}
%
\begin{barticle}[mr]
\bauthor{\bsnm{Kadane},~\bfnm{Joseph~B.}\binits{J.~B.}} \AND
\bauthor{\bsnm{Woodworth},~\bfnm{George~G.}\binits{G.~G.}}
(\byear{2004}).
\btitle{Hierarchical models for employment decisions}.
\bjournal{J. Bus. Econom. Statist.}
\bvolume{22}
\bpages{182--193}.
\bid{doi={10.1198/073500104000000073}, issn={0735-0015}, mr={2049920}}
\bptok{imsref}%
\end{barticle}
%
\endbibitem


\bibitem[\protect\citeauthoryear{Lin}{1997}]{Lin97}
%
\begin{barticle}[auto:STB|2013/05/03|14:24:43]
\bauthor{\bsnm{Lin},~\bfnm{D.~Y.}\binits{D.~Y.}}
(\byear{1997}).
\btitle{Non-parametric inference for cumulative incidence function in competing
risks studies}.
\bjournal{Stat. Med.}
\bvolume{16}
\bpages{901--910}.
\bptok{imsref}%
\end{barticle}
%
\endbibitem

\bibitem[\protect\citeauthoryear{Pan and Gastwirth}{2013}]{PanGas}
%
\begin{bmisc}[auto:STB|2013/05/03|14:24:43]
\bauthor{\bsnm{Pan},~\bfnm{Q.}\binits{Q.}} \AND
\bauthor{\bsnm{Gastwirth},~\bfnm{J.~L.}\binits{J.~L.}}
(\byear{2013}).
\bhowpublished{The appropriateness of survival analysis for determining
lost pay in discrimination cases when the number of plaintiffs exceeds the
number of job openings: Application of the ``Lost Chance'' doctrine to
Alexander V. Milwaukee. \textit{Law, Probability \& Risk} \textbf{12} 13--35}.
\bptok{imsref}%
\end{bmisc}
%
\endbibitem

\bibitem[\protect\citeauthoryear{Pan and Gastwirth}{2009}]{PanGas09}
%
\begin{barticle}[auto:STB|2013/05/03|14:24:43]
\bauthor{\bsnm{Pan},~\bfnm{Q.}\binits{Q.}} \AND
\bauthor{\bsnm{Gastwirth},~\bfnm{J.~L.}\binits{J.~L.}}
(\byear{2009}).
\btitle{Issues in the use of survival analysis to estimate damages in equal
employment cases}.
\bjournal{Law, Probability \& Risk}
\bvolume{8}
\bpages{1--24}.
\bptok{imsref}%
\end{barticle}
%
\endbibitem

\bibitem[\protect\citeauthoryear{Peng and Fine}{2007}]{PenFin07}
%
\begin{barticle}[mr]
\bauthor{\bsnm{Peng},~\bfnm{Limin}\binits{L.}} \AND
\bauthor{\bsnm{Fine},~\bfnm{Jason~P.}\binits{J.~P.}}
(\byear{2007}).
\btitle{Regression modeling of semicompeting risks data}.
\bjournal{Biometrics}
\bvolume{63}
\bpages{96--108, 311}.
\bid{doi={10.1111/j.1541-0420.2006.00621.x}, issn={0006-341X}, mr={2345579}}
\bptok{imsref}%
\end{barticle}
%
\endbibitem

\bibitem[\protect\citeauthoryear{Pollard}{1990}]{Pol90}
%
\begin{bbook}[mr]
\bauthor{\bsnm{Pollard},~\bfnm{David}\binits{D.}}
(\byear{1990}).
\btitle{Empirical Processes: Theory and Applications}.
\bseries{NSF-CBMS Regional Conference Series in Probability and Statistics}
\bvolume{2}.
\bpublisher{IMS}, \blocation{Hayward, CA}.
\bid{mr={1089429}}
\bptok{imsref}%
\end{bbook}
%
\endbibitem

\bibitem[\protect\citeauthoryear{Rubin}{1974}]{Rub74}
%
\begin{barticle}[auto:STB|2013/05/03|14:24:43]
\bauthor{\bsnm{Rubin},~\bfnm{D.~B.}\binits{D.~B.}}
(\byear{1974}).
\btitle{Estimating causal effects of treatments in randomized and nonrandomized
studies}.
\bjournal{J. of Educ. Psychol.}
\bvolume{66}
\bpages{688--701}.
\bptok{imsref}%
\end{barticle}
%
\endbibitem

\bibitem[\protect\citeauthoryear{Sen and Singer}{1993}]{SenSin93}
%
\begin{bbook}[mr]
\bauthor{\bsnm{Sen},~\bfnm{Pranab~K.}\binits{P.~K.}} \AND
\bauthor{\bsnm{Singer},~\bfnm{Julio~M.}\binits{J.~M.}}
(\byear{1993}).
\btitle{Large Sample Methods in Statistics: An Introduction with Applications}.
\bpublisher{Chapman \& Hall}, \blocation{New York}.
\bid{mr={1293125}}
\bptok{imsref}%
\end{bbook}
%
\endbibitem

\bibitem[\protect\citeauthoryear{Tableman and Stahel}{2009}]{TabSta09}
%
\begin{barticle}[mr]
\bauthor{\bsnm{Tableman},~\bfnm{Mara}\binits{M.}} \AND
\bauthor{\bsnm{Stahel},~\bfnm{Werner~A.}\binits{W.~A.}}
(\byear{2009}).
\btitle{Nonparametric methods for employment termination times with competing
causes}.
\bjournal{Stat. Interface}
\bvolume{2}
\bpages{37--44}.
\bid{issn={1938-7989}, mr={2500766}}
\bptok{imsref}%
\end{barticle}
%
\endbibitem
%

\bibitem[\protect\citeauthoryear{van~der Vaart and Wellner}{1996}]{vanWel96}
%
\begin{bbook}[mr]
\bauthor{\bparticle{van~der} \bsnm{Vaart},~\bfnm{Aad~W.}\binits{A.~W.}}
\AND
\bauthor{\bsnm{Wellner},~\bfnm{Jon~A.}\binits{J.~A.}}
(\byear{1996}).
\btitle{Weak Convergence and Empirical Processes: With Applications to
Statistics}.
\bpublisher{Springer}, \blocation{New York}.
\bid{mr={1385671}}
\bptok{imsref}%
\end{bbook}
%
\endbibitem

\bibitem[\protect\citeauthoryear{Woodworth and Kadane}{2010}]{WooKad10}
%
\begin{barticle}[mr]
\bauthor{\bsnm{Woodworth},~\bfnm{George}\binits{G.}} \AND
\bauthor{\bsnm{Kadane},~\bfnm{Joseph}\binits{J.}}
(\byear{2010}).
\btitle{Age- and time-varying proportional hazards models for employment
discrimination}.
\bjournal{Ann. Appl. Stat.}
\bvolume{4}
\bpages{1139--1157}.
\bid{doi={10.1214/10-AOAS330}, issn={1932-6157}, mr={2751336}}
\bptok{imsref}%
\end{barticle}
%
\endbibitem

\bibitem[\protect\citeauthoryear{Xu, Kalbfleisch and Tai}{2010}]{XuKalTai10}
%
\begin{barticle}[mr]
\bauthor{\bsnm{Xu},~\bfnm{Jinfeng}\binits{J.}},
\bauthor{\bsnm{Kalbfleisch},~\bfnm{John~D.}\binits{J.~D.}} \AND
\bauthor{\bsnm{Tai},~\bfnm{Beechoo}\binits{B.}}
(\byear{2010}).
\btitle{Statistical analysis of illness-death processes and
semicompeting risks
data}.
\bjournal{Biometrics}
\bvolume{66}
\bpages{716--725}.
\bid{doi={10.1111/j.1541-0420.2009.01340.x}, issn={0006-341X}, mr={2758207}}
\bptok{imsref}%
\end{barticle}
%
\endbibitem

\bibitem[\protect\citeauthoryear{Ye, Kalbfleisch and
Schaubel}{2007}]{YeKalSch07}
%
\begin{barticle}[mr]
\bauthor{\bsnm{Ye},~\bfnm{Yining}\binits{Y.}},
\bauthor{\bsnm{Kalbfleisch},~\bfnm{John~D.}\binits{J.~D.}} \AND
\bauthor{\bsnm{Schaubel},~\bfnm{Douglas~E.}\binits{D.~E.}}
(\byear{2007}).
\btitle{Semiparametric analysis of correlated recurrent and terminal events}.
\bjournal{Biometrics}
\bvolume{63}
\bpages{78--87, 311}.
\bid{doi={10.1111/j.1541-0420.2006.00677.x}, issn={0006-341X}, mr={2345577}}
\bptok{imsref}%
\end{barticle}
%
\endbibitem

\bibitem[\protect\citeauthoryear{Zhang and Schaubel}{2011}]{ZhaSch11}
%
\begin{barticle}[mr]
\bauthor{\bsnm{Zhang},~\bfnm{Min}\binits{M.}} \AND
\bauthor{\bsnm{Schaubel},~\bfnm{Douglas~E.}\binits{D.~E.}}
(\byear{2011}).
\btitle{Estimating differences in restricted mean lifetime using observational
data subject to dependent censoring}.
\bjournal{Biometrics}
\bvolume{67}
\bpages{740--749}.
\bid{doi={10.1111/j.1541-0420.2010.01503.x}, issn={0006-341X}, mr={2829128}}
\bptok{imsref}%
\end{barticle}
%
\endbibitem

\end{thebibliography}
\end{document}